\documentclass[11pt,a4paper]{article}

\usepackage[
top    = 2.50cm,
bottom = 2.50cm,
left   = 2.50cm,
right  = 2.50cm]{geometry}
\usepackage{jheppub}
\usepackage[sort&compress]{natbib}

\usepackage{amssymb}
\usepackage{graphicx}
\usepackage{slashed}
\usepackage{amsmath}
\usepackage{hyperref}
\usepackage{verbatim}
\usepackage{tabulary}
\usepackage{booktabs}
\usepackage{tensor}
 \usepackage{multirow}
 \usepackage{enumerate} 
 \numberwithin{equation}{subsection}

\newcommand{\eql}[2]{\begin{equation}\label{#1}{\begin{split}#2\end{split}}\end{equation}}
\newcommand{\eq}[2][ ]{\begin{equation}\label{#1}{\begin{split}#2\end{split}}\end{equation}}

\newcommand{\gmn}{g^{\mu\nu}}

\newcommand{\al}{\alpha}
\newcommand{\be}{\beta}
\newcommand{\te}{\theta}
\newcommand{\la}{\lambda}

\newcommand{\dm}{\partial_\mu}
\newcommand{\dn}{\partial_\nu}

\newcommand{\dr}{\partial_r}
\newcommand{\dt}{\partial_t}
\newcommand{\dx}{\partial_x}

\newcommand{\dth}{\partial_\theta}

\newcommand{\zb}{\bar{z}}

\newcommand{\onov}[1]{\frac{1}{#1}}
\newcommand{\mat}[1]{\left(\begin{matrix} #1 \end{matrix}\right)}

\newcommand{\lag}{\mathcal{L}}

\newcommand{\zeb}{\bar{\zeta}}
\newcommand{\qq}{\mathcal{Q}}
\renewcommand{\arraystretch}{1.5}

\title{Supersymmetric Field Theories on $AdS_p\times S^q$}

\preprint{WIS/15/09-NOV-DPPA}

\author{
Ofer Aharony, Micha Berkooz, Avner Karasik and Talya Vaknin
\\
\it{Department of Particle Physics and Astrophysics,\\
Weizmann Institute of Science, Rehovot 7610001, Israel}
}
\emailAdd{Ofer.Aharony@weizmann.ac.il}
\emailAdd{Micha.Berkooz@weizmann.ac.il}
\emailAdd{Avner.Karasik@weizmann.ac.il}
\emailAdd{Talya.Vaknin@weizmann.ac.il}

\abstract{
In this paper we study supersymmetric field theories on an $AdS_p\times S^q$ space-time that preserves their full supersymmetry. This is an interesting example of supersymmetry on a non-compact curved space. The supersymmetry algebra on such a space is a $(p-1)$-dimensional superconformal algebra, and we classify all possible algebras that can arise for $p \geq 3$. In some $AdS_3$ cases more than one superconformal algebra can arise from the same field theory. We discuss in detail the special case of four dimensional field theories with ${\cal N}=1$ and ${\cal N}=2$ supersymmetry on $AdS_3\times S^1$.
}

\begin{document}
\maketitle
\bibliographystyle{JHEP}

\section{Introduction}

In the last few years, it has been useful to study
supersymmetric field theories on curved manifolds. In many cases exact results in such backgrounds (when they preserve supersymmetry) may be computed (following \cite{Pestun:2007rz}), including partition functions and some expectation values. In some cases these contain information beyond the usual exactly-computable results in flat space. This allows us to learn more about these field theories, even when they are strongly coupled, providing new windows into strongly coupled field theories.

So far, this study has been almost exclusively limited to compact curved manifolds. In this paper we take first steps towards a systematic study of supersymmetric field theories on non-compact curved space-times, by looking at the specific example of  $d$-dimensional supersymmetric (SUSY) field theories on $AdS_{d-q}\times S^q$. This example is maximally symmetric, and, for an appropriate definition of the theory and of the boundary conditions, can often preserve the full supersymmetry of the $d$-dimensional field theory. Another reason for being interested in this specific example is that  field theories on $AdS_{d-q}\times S^q$ often arise on various branes and singularities in string theories on anti-de Sitter ($AdS$) space, and in some cases they can be decoupled from the full gravitational theory. In a specific example, the $6d$ $\mathcal{N}=(2,0)$ superconformal field theory on $AdS_5\times S^1$ was recently embedded in string theory \cite{Aharony:2015zea}, and this led to surprising results. Four dimensional supersymmetric theories on $AdS_4$ were studied in detail in the past, for instance in \cite{Zumino:1977av,Ivanov:1979ft,Ivanov:1980vb,Sakai:1984nc,Burgess:1984rz,Burgess:1984ti,Burges:1985qq,Adams:2011vw,Aharony:2010ay,Aharony:2012jf} (this example is particularly interesting since it is related by a conformal transformation to four dimensional theories on a half-space). Supersymmetric theories on $AdS_{3,4,5}$ were studied in \cite{Kuzenko:2007aj,Kuzenko:2008kw,Kuzenko:2008qw,Kuzenko:2011rd,Butter:2012jj,Kuzenko:2012bc,Butter:2012vm,Kuzenko:2013uya,Kuzenko:2014mva,Kuzenko:2014eqa}, on $AdS_5$ in \cite{Shuster:1999zf}, and on $AdS_d$ and $AdS_p\times S^p$ in \cite{Bandos:2002nn}. Other examples did not attract much attention. Another example worth mentioning is  the $\mathcal N=4$ super-Yang-Mills theory on $AdS_3\times S^1$, that can be embedded in string theory by considering D3-branes on $AdS_3\times S^1$ in type IIB string theory on $AdS_5\times S^5$.

We limit our study to anti-de Sitter spaces of at least three dimensions.
In section \ref{SCalgebra} we go over all possible UV-complete supersymmetric field theories on $AdS_{d-q}\times S^q$, and classify the possible supersymmetry algebras that they can have that preserve the full supersymmetry. These algebras always contain the isometries of $AdS_{d-q}$, so if we start from a field theory with $n$ supercharges, we obtain a superconformal algebra in $(d-q-1)$ dimensions with $n$ total supercharges, $n/2$ regular and $n/2$ conformal supercharges.
When $(d-q)>3$ such a superconformal algebra is unique, so the only questions are whether we can preserve all the supercharges and obtain this algebra or not. For $(d-q)=3$ there are several $2d$ superconformal algebras with the same total number of supercharges, and we often find that more than one possibility can be realized by the same field theory when we put it on $AdS_{d-q}\times S^q$.

There are several different methods to perform this analysis. One
possibility is to use the formalism of supersymmetry in curved space,
by coupling the supersymmetric theory to a background supergravity
theory with an appropriate metric (and additional background fields).
This formalism is only available for some cases, but when it is
available it is straightforward to find that the $AdS_{d-q}
\times S^q$ background can preserve all the supercharges,
and to write the supersymmetric actions
and Killing spinor equations on $AdS_{d-q}\times S^q$. There are
actually two variations of this formalism. One can use a ``regular''
supergravity containing background fields coupled to the usual
supersymmetry algebra, or one can use a ``conformal'' supergravity
containing background fields coupling to the superconformal algebra
(see, for instance, \cite{Kuzenko:2007aj,Kuzenko:2008kw,Kuzenko:2008qw,Kuzenko:2011rd,Butter:2012jj,Kuzenko:2012bc,Butter:2012vm,Kuzenko:2013uya,Kuzenko:2014mva,Kuzenko:2014eqa}).
When both formalisms are available they are identical, since the
``regular'' supergravity arises as a particular gauge-fixing of the
``conformal'' supergravity. In particular they give rise to the
same Killing spinor equations, though they may be written in
terms of different background fields. In section \ref{AdS3S1} we use a ``regular''
supergravity to analyze (following \cite{Festuccia:2011ws}) the case of
$4d$ $\mathcal{N}=1$ theories on $AdS_3\times S^1$, and we use
``conformal'' supergravity to analyze the case of $4d$ $\mathcal{N}=2$
theories on the same space. We analyze these two cases in
detail, constructing explicitly the Killing spinors and the
supersymmetry transformations. For various different
free $4d$ theories, we analyze in detail the spectrum that
we obtain on $AdS_3\times S^1$, and the corresponding
representations of the $2d$ superconformal algebra.

In section \ref{SCalgebra}, in order to perform the complete classification, we use
a completely different method. This is a more general approach,
that does not require knowledge of the precise background supergravity that is relevant
(and that is not always available); this method was previously used in \cite{Bandos:2002nn}. To do this we note that
$AdS_{d-q}\times S^q$ (with equal radii for $AdS$ and the sphere) is related to flat space $R^d$ by a conformal
transformation. Thus, the resulting $(d-q-1)$-dimensional
superconformal algebra must be a subalgebra of the
$d$-dimensional superconformal algebra, and we can just
classify all such subalgebras (that contain half of the fermionic
charges of the $d$-dimensional superconformal algebra, and
the isometries of $AdS_{d-q}\times S^q$). In this method
the supercharges that we preserve on $AdS_{d-q}\times S^q$ are
combinations of regular supercharges and conformal
supercharges, that arise after performing the conformal
transformation from flat space. This is similar to what we
obtain by coupling our theory to conformal supergravity,
but here we do not need to use any details of this coupling,
and it is clear from the discussion above that the results
apply to general supersymmetric field theories on $AdS_{d-q}\times S^q$
(not necessarily superconformal). The embedding of the $(d-q-1)$-dimensional
superconformal algebra into the $d$-dimensional superconformal algebra
immediately tells us which $d$-dimensional R-symmetries are required for
preserving supersymmetry on $AdS_{d-q}\times S^q$. Our analysis is limited to
supersymmetric field theories that have a UV-completion
as superconformal field theories; we do not discuss in this
paper theories with no known field-theoretic UV completion,
such as the $6d$ $\mathcal{N}=(1,1)$ supersymmetric Yang-Mills theory.

It is important to emphasize that a supersymmetric field theory on $AdS_{d-q}\times S^q$ is not equivalent to a $(d-q-1)$-dimensional superconformal theory; for instance it does not contain a graviton that would map to the energy-momentum tensor of such a theory. However, such theories do have a natural action of the $(d-q-1)$-dimensional superconformal algebra, such that their states and fields sit in representations of this algebra, and they can arise as decoupled subsectors of full-fledged 
$(d-q-1)$-dimensional superconformal theories \cite{Aharony:2015zea}.

One new aspect which arises for theories on non-compact space-times like $AdS$ is the need to specify boundary conditions, in particular in a way that preserves the full supersymmetry. We do not discuss this issue in general here, assuming that such a choice is always possible. In the cases that we discuss in detail we explicitly discuss some boundary conditions which preserve supersymmetry. Often there are many different choices of maximally supersymmetric boundary conditions, in particular for non-Abelian gauge theories \cite{Gaiotto:2008sa}.

It would be interesting to understand what are the specific quantities that can be computed exactly for supersymmetric field theories on $AdS_{d-q}\times S^q$, by localization or other methods. We leave this to future work.
 
\section{Superconformal field theories on $AdS_{d-q}\times S^q$}
\label{SCalgebra}

In this paper we study supersymmetric field theories in $d=3,4,5,6$, that are put on manifolds  of the form $AdS_p\times S^q$ ($d=p+q$, $p\geq 3$) in a way that preserves the full supersymmetry (SUSY). While some partial results are available (for example, for $4d$ $\mathcal{N}=1$ theories \cite{Dumitrescu:2012ha}), a general method for analyzing and constructing supersymmetric theories on curved space is not yet available; in some cases one can use a coupling to background supergravity fields for this. However, on space-times that include anti-de Sitter space, we have the advantage that any supersymmetry algebra must include the isometry algebra $SO(p-1,2)$, which means that it must be equivalent to a superconformal algebra in $(p-1)$ dimensions, with the same total number of supercharges. 

We can use the following trick to analyze all possible SUSY algebras that can arise from $(p+q)$-dimensional supersymmetric theories on $AdS_p\times S^q$.
When the $AdS$ space and the sphere have equal radii of curvature, the space $AdS_p\times S^q$ is conformally equivalent to $(p+q)$-dimensional flat space. We can use for $AdS_p\times S^q$ the metric
\eq{ds^2 = L^2 \left[ \frac{dz^2 + dx^{\mu} dx_{\mu}}{z^2} + d\Omega_q^2 \right], }
where $\mu = 0,\cdots,p-1$, $d\Omega_q^2$ is the metric on a unit $S^q$, and the boundary is at $z=0$. Then, multiplying the metric by $z^2 / L^2$ gives the metric on flat space, where $z$ is a radial coordinate in $(q+1)$ dimensions. Boundary conditions imposed on $AdS$ space may lead to singularities of various fields on the $(p-1)$-dimensional subspace $z=0$. 

Suppose we have a $(p+q)$-dimensional superconformal theory, whose symmetry algebra includes $n$ supercharges and also $n$ superconformal charges. 
The fact that $AdS_p\times S^q$ is related to flat space by a conformal transformation, means that the supersymmetry algebra on $AdS_p\times S^q$ must be a subalgebra of the superconformal algebra in $(p+q)$-dimensions. It is clear that the boundary conditions on $AdS_p$ must break at least half of the total number of fermionic generators of the superconformal theory. We will be interested in the cases where exactly half of the fermionic generators are broken, leading to a supersymmetry algebra on $AdS_p\times S^q$ with $n$ fermionic generators. Thus, our goal will be to classify all possible $(p-1)$ dimensional superconformal algebras with $n$ total supercharges ($n/2$ regular supercharges and $n/2$ superconformal charges in $(p-1)$ dimensions) that arise as subalgebras of $(p+q)$-dimensional superconformal algebras with $2n$ total supercharges. The superconformal groups in different dimensions on flat space were classified by Nahm \cite{Nahm:1977tg}, and we will use this classification in our analysis. 

Naively this classification is only relevant for superconformal theories in $(p+q)$ dimensions. However, it is clear that if we have a general supersymmetric theory on $AdS_p\times S^q$, it cannot preserve a larger supersymmetry than that of a superconformal theory on the same space. Thus, the same classification will give us all supersymmetry algebras that can arise when we put a general UV-complete supersymmetric theory on $AdS_p\times S^q$ in a way that preserves the same number of supercharges as we had in $(p+q)$ dimensions. Similarly, it is clear that when we discuss $AdS_p$ and $S^q$ spaces with different radii of curvature, the supersymmetry algebra cannot be larger than the one which arises for equal radii, so any supersymmetry algebra that arises for such space-times must also be included in our classification. For $q=1$ one can show that
the same supersymmetry algebra arises for any ratio of the radii.

In this section we will describe this classification of all possible symmetry algebras that preserve all supercharges of a $(p+q)$-dimensional theory on $AdS_p\times S^q$, going over all possible $p \geq 3$ cases one by one. The results are summarized in appendix \ref{Summary}.

\subsection{Four dimensional theories on $AdS_3\times S^1$}

In four dimensions we have four consistent superconformal algebras that do not contain higher spin conserved currents : $\mathcal{N}=1$, $\mathcal{N}=2$, $\mathcal{N}=3$ and $\mathcal{N}=4$. Each algebra contains $4\mathcal{N}$ regular supercharges, and $4\mathcal{N}$ superconformal charges. We would like to analyze the possible algebras that can arise for supersymmetric theories on $AdS_3\times S^1$, preserving all $4\mathcal{N}$ supercharges. As discussed above, such algebras should be two dimensional superconformal algebras with $2\mathcal{N}$ regular supercharges, that arise as subalgebras of the four dimensional superconformal algebras. Thus, they must be $(p,2\mathcal{N}-p)$ superconformal algebras in two dimensions. Throughout this section we will follow the notations of \cite{Freedman:2012zz}; our conventions may be found in appendix \ref{conventions}.

\subsubsection{$\mathcal{N}=1$}
\label{d4N1}

In this section we describe the algebraic structure of these theories; the construction of some explicit field theories for this case will be discussed in detail in section \ref{AdS3S1}.

The $4d$ $\mathcal{N}=1$ superconformal algebra includes the fermionic generators $Q_{\alpha}$ and $S_\beta$, which we take to be Majorana spinors with $\alpha,\beta=1,2,3,4$, and the bosonic generators: Lorentz transformations $M_{\mu\nu}$ of $SO(3,1)$, translations $P_\mu$, special conformal transformations $K_\mu$ and dilatations $D$, with $\mu,\nu=0,\cdots,3$. Moreover, it includes a $U(1)$ R-symmetry whose generator we denote by $T$. The bosonic generators satisfy the $SO(4,2)$ algebra, and together with the fermionic generators they satisfy the $SU(2,2|1)$ algebra:

\eq{&\{Q_\al,Q^\be\}=-\onov{2}(\gamma^\mu)_\al^\be P_\mu\quad ,\quad \{S_\al,S^\be\}=-\onov{2}(\gamma^\mu)_\al^\be K_\mu\quad, \\
&\{Q_\al,S^\be\}=-\onov{2}\delta_\al^\be D-\onov{4}(\gamma^{\mu\nu})_\al^\be M_{\mu\nu}+\frac{i}{2}(\gamma^*)_\al^\be T\quad,\\
&[M_{\mu\nu},Q]=-\onov{2}\gamma_{\mu\nu}Q,\quad [M_{\mu\nu},S]=-\onov{2}\gamma_{\mu\nu}S\quad,\\
&[T,Q]=-\frac{3i}{2}\gamma_*Q\quad ,\quad [T,S]=\frac{3i}{2}\gamma_*S\quad,\\
&[K_\mu,Q]=\gamma_\mu S\quad,\quad [P_\mu,S]=\gamma_\mu Q\quad.
}
Here $\gamma_*$ is the chiral gamma matrix, that exists in any even number of dimensions.

In order to have a consistent superalgebra on $AdS_3\times S^1$, we need a subalgebra of this superconformal algebra that includes the $AdS_3\times S^1$ isometries, $SO(2,2)\times SO(2) \subset SO(4,2)$, and it should include half of the fermionic operators $Q$ and $S$, that must close on this subalgebra. We expect the $SO(2)$ generator of rotations on the circle to include $P_3$, and to commute with all other bosonic generators, in particular with the $SO(2,2)$ generators. The correct choice turns out to be $P_3 - c^2 K_3$, where we show in appendix \ref{KSE} that the absolute value of $c$ is related to the curvature of $AdS_3$ by \eq{c^2=\onov{24}\mathcal{R}.}

The generator $P_3 - c^2 K_3$ does not commute with the following eight bosonic generators: $P_a-c^2K_a$, $M_{a3}$, $D$, and $P_3+c^2K_3$ ($a=0,1,2$), so these generators will not be part of the resulting supersymmetry algebra. Six of the remaining bosonic generators, $M_{ab}$ and $P_a+c^2K_a$, form an $SO(2,2)$ algebra.
In order to see this it is convenient to define the operators
\eql{ladef}{P_a+c^2K_a-c\epsilon^{bc}_aM_{bc}=\mat{L_0\\L_1\\L_2}=\mat{P_0+c^2K_0+2cM_{12}\\P_1+c^2K_1-2cM_{20}\\P_2+c^2K_2-2cM_{01}}\quad,}
and 
\eq{P_a+c^2K_a+c\epsilon^{bc}_aM_{bc}=\mat{\bar L_0\\\bar L_1\\\bar L_2}=\mat{P_0+c^2K_0-2cM_{12}\\P_1+c^2K_1+2cM_{20}\\P_2+c^2K_2+2cM_{01}}\quad.}
The $SO(2,2) \simeq SO(2,1)\times SO(2,1)$ algebra may then be written as:
\eq{&[L_1,L_2]=4cL_0\quad,\quad [L_0,L_1]=-4cL_2\quad,\quad [L_2,L_0]=-4cL_1\quad,\\
&[\bar{L}_1,\bar{L}_2]=-4c\bar{L}_0\quad,\quad [\bar L_0,\bar L_1]=4c\bar{L}_2\quad,\quad [\bar{L}_2,\bar{L}_0]=4c\bar{L}_1\quad,\\
&[L_a, \bar{L}_b] = 0.}
When $c>0$, the $L_a$ are the right-moving $SO(2,1)$ charges and the ${\bar L}_a$ are the left-moving $SO(2,1)$ charges. Without loss of generality we will assume from here on that $c>0$; the other sign is related to this by a two dimensional parity transformation.
When the $SO(2,2)$ is embedded into a full Virasoro algebra (which may or may not be the case), these charges are proportional to the ${\cal L}_{-1,0,1}$ and $\bar{\cal L}_{-1,0,1}$ charges of the Virasoro algebra (the constant $c$ above should not be confused with the central charge of the Virasoro algebra, which can be non-zero when our field theory is embedded into some gravitational theory).

The supercharges that close only on the remaining bosonic generators mentioned above are half of all the fermionic generators, that may be written as $\qq\equiv Q+ic\gamma^*\gamma_3S$. Their algebra is
\eq{\{\qq_\al,\qq^\be\}=&-\onov{2}\tensor{(\gamma^3)}{_\al^\be} (P_3-c^2K_3)-\onov{2}\tensor{(\gamma^a)}{_\al^\be}(P_a+c^2K_a)-c\tensor{(\gamma^3)}{_\al^\be}T+\frac{ic}{2}\tensor{(\gamma^{ab}\gamma^3\gamma^*)}{_\al^\be} M_{ab}\quad,\\=&-\onov{2}\tensor{(\gamma^3)}{_\alpha^\be}(P_3-c^2K_3+2cT)-\onov{2}\tensor{(\gamma^a)}{_\al^\be}L_a\quad.}
We can identify this as part of the two dimensional right-moving ${\cal N}=2$ superconformal algebra, where we identify $(P_3 - c^2 K_3 + 2 c T)$ as the two dimensional R-charge; note that this is a mixture of the isometry of the circle and the $4d$ R-charge. Together with the other generators written above we have the ${\cal N}=(0,2)$ superconformal algebra (when $c$ is positive), with a bosonic subgroup $ SO(2,2)\times SO(2)$; in particular
\eq{&[L_a,\qq]=2ic\gamma^*\gamma^3\gamma_a\qq,\qquad\qquad
[\bar{L}_a,\qq]=0\quad,\\&[P_3-c^2K_3+2cT,\qq]=-2ic\gamma^*\qq\quad.}

In addition to this algebra we have an extra $U(1)$ generator $(P_3 - c^2 K_3 - 2 c T)$ that commutes with all the generators of the superconformal algebra. When this generator is preserved, it gives an extra global $U(1)$ symmetry, but we can also preserve the same amount of supersymmetry on $AdS_3\times S^1$ when this generator is broken, and only the specific combination $(P_3 - c^2 K_3 + 2 c T)$ is conserved; we will see examples of both possibilities in section \ref{AdS3S1}. However, the fact that we must preserve the combination $(P_3 - c^2 K_3 + 2 c T)$ which appears on the right-hand side of the supersymmetry algebra on $AdS_3\times S^1$ means that even when the $4d$ theory that we start with is not superconformal, it must still have an R-symmetry in order to preserve supersymmetry on $AdS_3\times S^1$ \cite{Festuccia:2011ws,Dumitrescu:2012ha}. This will be true also in our subsequent examples -- preserving supersymmetry on $AdS_p\times S^q$ requires having in the $(p+q)$-dimensional field theory all the R-symmetries that appear on the right-hand side of the supersymmetry algebra on $AdS_p\times S^q$.

Note that even though just the counting of supercharges would have allowed also a $2d$ ${\cal N}=(1,1)$ superconformal algebra in this case, we see that this possibility does not arise. We will see also in the other cases below that on $AdS_3\times S^q$ ($q \geq 1$) we always obtain an even number of left-moving and of right-moving supersymmetry generators.

\subsubsection{$\mathcal{N}=2$}
\label{d4N2}
In the $4d$ $\mathcal{N}=2$ case, we use a convention where we have eight complex Weyl supercharges $Q_{i\alpha}$ and $S^j_\beta$, with $i,j=1,2$, $\alpha,\beta=1,2,3,4$, and a $SU(2)\times U(1)$ R-symmetry generated by a traceless $2\times 2$ matrix $U_i^j$ and by $T$. The position of the $R$ symmetry index $i,j,\cdots$ is used to distinguish between left-handed and right-handed supercharges in the following way:
 \eql{Chiral}{\gamma^*Q_{i\al}=Q_{i\al}\ ,\ \gamma^*Q^i_{\al}=-Q^i_{\al}\ ,\ \gamma^*S^i_{\al}=S^i_{\al}\ ,\ \gamma^*S_{i\al}=-S_{i\al}.}
 In order to connect the results here to our later results in $5d$ and $6d$, we follow \cite{Freedman:2012zz} and treat the Weyl spinors as four component spinors, with two components vanishing according to (\ref{Chiral}). The spinor indices $\alpha,\beta,\cdots$ can be raised and lowered by the charge conjugation matrix $C^{\alpha\beta}$. In the usual notations of \cite{Wess:1992cp}, our $Q_i$'s are the $Q$'s, and our $Q^i$'s are the $\overline{Q}$'s.

 Together with the other bosonic generators these satisfy the $SU(2,2|2)$ superconformal algebra
\eq{&\{Q_{i\al},Q^{j\be}\}=-\onov{2}\delta_i^j(\gamma^\mu)_\al^\be P_\mu\quad,\quad \{S_{i\al},S^{j\be}\}=-\onov{2}\delta_i^j(\gamma^\mu)_\al^\be K_\mu\quad,\\
&\{Q_{\al}^i,Q^{\be}_j\}=-\onov{2}\delta_j^i(\gamma^\mu)_\al^\be P_\mu\quad,\quad \{S_{\al}^i,S^{\be}_j\}=-\onov{2}\delta_j^i(\gamma^\mu)_\al^\be K_\mu\quad,\\
&\{Q_{i\al},S^{j\be}\}=-\onov{2}\delta_i^j\delta_\al^\be D-\onov{4}\delta_i^j(\gamma^{\mu\nu})_\al^\be M_{\mu\nu}+\frac{i}{2}\delta_i^j\delta_\al^\be T-\delta_\al^\be U_i^j\quad,\\
&\{Q_{\al}^i,S^{\be}_j\}=-\onov{2}\delta_j^i\delta_\al^\be D-\onov{4}\delta_j^i(\gamma^{\mu\nu})_\al^\be M_{\mu\nu}-\frac{i}{2}\delta_j^i\delta_\al^\be T+\delta_\al^\be U_j^i\quad,\\
&[U_i^j,Q^k]=\delta_i^kQ^j-\onov{2}\delta_i^jQ^k\quad,\quad [U_i^j,S^k]=\delta_i^k S^j-\onov{2}\delta_i^jS^k\quad,\\
&[U_i^j,Q_k]=-\delta_k^jQ_i+\onov{2}\delta_i^jQ_k\quad,\quad [U_i^j,S_k]=-\delta_k^j S_i+\onov{2}\delta_i^jS_k\quad,\\
&[T,Q^i]=\frac{i}{2}Q^i\quad,\quad [T,S^i]=\frac{i}{2}S^i\quad,\\
&[T,Q_i]=-\frac{i}{2}Q_i\quad,\quad [T,S_i]=-\frac{i}{2}S_i\quad.
}

As in the $\mathcal{N}=1$ case, we choose a subgroup of the bosonic generators to give the isometry algebra of $AdS_3\times S^1$, and we correlate the ordinary supercharges and the superconformal charges so that we get a consistent superalgebra on $AdS_3\times S^1$. In the $\mathcal{N}=2$ case there are two different ways to do so, which are a straightforward generalization of the $\mathcal{N}=1$ case (see also \cite{Butter:2015tra}). 

The first option is the diagonal option preserving $SU(2)_R$, where for $i=1,2$ we choose the conserved fermionic generators to be 
\eq{&\qq_i\equiv Q_i+ic\gamma^*\gamma^3S_i=Q_i+ic\gamma^3S_i\quad,\\
&\qq^i\equiv Q^i+ic\gamma^*\gamma^3S^i=Q^i-ic\gamma^3S^i\quad.}
These satisfy the algebra
\eql{firstop}{&\{\qq_i,\qq_j\}=\{\qq^i,\qq^j\}=0\quad,\\
&\{\qq_i,\qq^j\}=-\onov{2}\delta_i^j\gamma^3(P_3-c^2K_3)-\onov{2}\delta_i^j\gamma^a(P_a+c^2K_a)-\frac{ic}{2}\delta_i^j\gamma^{ab}\gamma^3M_{ab}-c\delta_i^j\gamma^3T-2ic\gamma^3U_i^j\quad,\\&[P_3-c^2K_3,\qq_i]=ic\qq_i\quad,\quad [P_3-c^2K_3,\qq^i]=-ic\qq^i\quad,\\
&[M_{ab},\qq_i]=-\onov{2}\gamma_{ab}\qq_i\quad,\quad [M_{ab},\qq^i]=-\onov{2}\gamma_{ab}\qq^i\quad,\\
&[P_a+c^2K_a,\qq_i]=ic\gamma^3\gamma_a\qq_i\quad,\quad[P_a+c^2K_a,\qq^i]=-ic\gamma^3\gamma_a\qq^i\quad,\\
&[T,\qq^i]=\frac{i}{2}\qq^i\quad,\quad[T,\qq_i]=-\frac{i}{2}\qq_i\quad,\\
&[U_i^j,\qq_k]=-\delta_k^j\qq_i+\onov{2}\delta_i^j\qq_k\quad,\\
&[U_i^j,\qq^k]=\delta_i^k\qq^j-\onov{2}\delta_i^j\qq^k\quad
.}
Notice that $P_3-c^2K_3+2cT$ appears on the right-hand side of $\{\qq_i,\qq^j\}$, but it commutes with all the other generators in the theory. Therefore, $k_\te \equiv P_3-c^2K_3+2cT$ is a central charge and fixed within a representation. This should not be confused with the central charge of the (super)-Virasoro algebra, which does not appear in its global subalgebra that we obtain here.
We can rewrite the second line of \eqref{firstop} in the following way:
\eq{&\{\qq_1,\qq^1\}=-\onov{2}\gamma^aL_a+2c\gamma^3U_3-\onov{2}\gamma^3k_\te\quad,\\
&\{\qq_2,\qq^2\}=-\onov{2}\gamma^aL_a-2c\gamma^3U_3-\onov{2}\gamma^3k_\te\quad,\\&\{\qq_1,\qq^2\}=2c\gamma^3(U_1-iU_2)\quad,\\ &\{\qq_2,\qq^1\}=2c\gamma^3(U_1+iU_2)\quad,
}
with $L_a$ defined as in \eqref{ladef}, and $U_i$ defined by
\eq{U_i^j=i(U_1\sigma_1+U_2\sigma_2+U_3\sigma_3)_i^j\quad.}
This supersymmetry algebra (together with the commutators of the bosonic generators) is isomorphic to the `small' $\mathcal{N}=(0,4)$ superconformal algebra in two dimensions, which contains an $SU(2)$ R-symmetry, together with an extra central charge $k_\te$. Note that this central charge is consistent with the algebra we wrote, but not with its extension to a super-Virasoro algebra; thus if we embed our theory into a theory that has this extended algebra, $k_\te$ must vanish.

The other possible option for $\mathcal{N}=2$ theories on $AdS_3\times S^1$  is  
\eq{&\qq_1\equiv Q_1+ic\gamma^3S_2\quad,\quad \qq_2\equiv Q_2+ic\gamma^3S_1\quad,\\
&\qq^1\equiv Q^1-ic\gamma^3S^2\quad,\quad \qq^2\equiv Q^2-ic\gamma^3S^1.}
It will be convenient to define the following basis for the charges
\eq{\qq_\pm\equiv \onov{\sqrt{2}}(\qq_1\pm\qq_2)\quad,\quad  \qq^\pm\equiv \onov{\sqrt{2}}(\qq^1\pm\qq^2)\quad.}
The algebra that includes this specific half of the supercharges is then
\eql{pm}{&\{\qq^\pm,\qq^\pm\}=\{\qq^\pm,\qq^\mp\}=\{\qq_\pm,\qq_\pm\}=\{\qq_\pm,\qq_\mp\}=\{\qq_\pm,\qq^\mp\}=0\quad,\\
&\{\qq_+,\qq^+\}=-\onov{2}\gamma^a(P_a+c^2K_a)-\frac{ic}{2}\gamma^{ab}\gamma^3M_{ab}-\onov{2}\gamma^3(P_3-c^2K_3)+2c\gamma^3U_1-c\gamma^3T\quad,\\
&\{\qq_-,\qq^-\}=-\onov{2}\gamma^a(P_a+c^2K_a)+\frac{ic}{2}\gamma^{ab}\gamma^3M_{ab}-\onov{2}\gamma^3(P_3-c^2K_3)+2c\gamma^3U_1+c\gamma^3T\quad,\\
&[P_3-c^2K_3,\qq_{\pm}]=\pm ic\qq_{\pm}\quad,\quad [P_3-c^2K_3,\qq^{\pm}]=\mp ic\qq^{\pm}\quad,\\
&[M_{ab},\qq_{\pm}]=-\frac{1}{2}\gamma_{ab}\qq_{\pm}\quad,\quad [M_{ab},\qq^{\pm}]=-\frac{1}{2}\gamma_{ab}\qq^{\pm}\quad,\\
&[P_a+c^2K_a,\qq^+]=ic\gamma^3\gamma_a\qq^+\quad,\quad [P_a+c^2K_a,\qq^-]=-ic\gamma^3\gamma_a\qq^-\quad,\\
&[P_a+c^2K_a,\qq_+]=-ic\gamma^3\gamma_a\qq_+\quad,\quad [P_a+c^2K_a,\qq_-]=ic\gamma^3\gamma_a\qq_-\quad,\\
&[T,\qq_{\pm}]=-\frac{i}{2}\qq_{\pm},\quad [T,\qq^{\pm}]=\frac{i}{2}\qq^{\pm},\quad [2U_1,\qq_{\pm}]=\pm i\qq_{\pm},\quad [2U_1,\qq^{\pm}]=\mp i\qq^{\pm}\quad.
}

Notice that  we get two separate subalgebras, each satisfying the $\mathcal{N}=1$ algebra of section \ref{d4N1}, but with opposite chirality.

These subalgebras contain two independent $U(1)$ generators 
\eql{tpm}{\mathcal{T}_\pm=\onov{2}(P_3-c^2K_3)-2cU_1\pm cT\quad,}
and each one of them acts on half of the supercharges: 
\eq{[\mathcal{T}_{\pm},\qq^\pm]=\pm ic\qq^\pm\quad,\quad [\mathcal{T}_{\pm},\qq^{\mp}]=0\quad,\quad
[\mathcal{T}_\pm,\qq_\pm]=\mp ic\qq_\pm\quad,\quad [\mathcal{T}_{\pm},\qq_{\mp}]=0\quad.}
This algebra is the $\mathcal{N}=(2,2)$ superconformal algebra in two dimensions, with left-moving and right-moving $U(1)_R$ generators ${\mathcal{T}_{\pm}}$.

\subsubsection{$\mathcal{N}=4$}
\label{d4N4}

In the $4d$ $\mathcal{N}=4$ case we have 16 complex Weyl supercharges satisfying the superconformal algebra $PSU(2,2|4)$. In a similar
notation to the previous subsection
\eq{&\{Q_{i\al},Q^{j\be}\}=-\onov{2}\delta_i^j(\gamma^\mu)_\al^\be P_\mu\quad,\quad \{S_{i\al},S^{j\be}\}=-\onov{2}\delta_i^j(\gamma^\mu)_\al^\be K_\mu\quad,\\
&\{Q_{\al}^i,Q^{\be}_j\}=-\onov{2}\delta_j^i(\gamma^\mu)_\al^\be P_\mu\quad,\quad \{S_{\al}^i,S^{\be}_j\}=-\onov{2}\delta_j^i(\gamma^\mu)_\al^\be K_\mu\quad,\\
&\{Q_{i\al},S^{j\be}\}=-\onov{2}\delta_i^j\delta_\al^\be D-\onov{4}\delta_i^j(\gamma^{\mu\nu})_\al^\be M_{\mu\nu}-\delta_\al^\be U_i^j\quad,\\
&\{Q_{\al}^i,S^{\be}_j\}=-\onov{2}\delta_j^i\delta_\al^\be D-\onov{4}\delta_j^i(\gamma^{\mu\nu})_\al^\be M_{\mu\nu}+\delta_\al^\be U_j^i\quad,\\
&[U_i^j,Q^k]=\delta_i^kQ^j-\onov{4}\delta_i^jQ^k\quad,\quad [U_i^j,S^k]=\delta_i^k S^j-\onov{4}\delta_i^jS^k\quad,\\
&[U_i^j,Q_k]=-\delta_k^jQ_i+\onov{4}\delta_i^jQ_k\quad,\quad [U_i^j,S_k]=-\delta_k^j S_i+\onov{4}\delta_i^jS_k\quad,\\
& [U_i^j,U_k^l]=\delta_i^lU_k^j-\delta_k^jU_i^l\quad,
}
where now $i,j=1,2,3,4$ are $SU(4)_R$ indices.

As in the previous cases, we define supercharges $\qq_{i i'}\equiv Q_i+ic\gamma^*\gamma^3S_{i'}$, where we begin with arbitrary and independent $i$ and $i'$. We wish to see which combinations of indices will close on the isometries of $AdS_3\times S^1$, chosen as above. Generally, the commutation relations of these supercharges are
\eq{\{\qq_{i i'},\qq^{j j'}\}=
&-\onov{2}\gamma^3(\delta_i^jP_3-c^2\delta_{i'}^{j'}K_3)-\onov{2}\gamma^a(\delta_i^jP_a+c^2\delta_{i'}^{j'}K_a)-\frac{ic}{4}\gamma^{ab}\gamma^3M_{ab}(\delta_i^{j'}+\delta_{i'}^j)\\&
-\frac{ic}{2}(\delta_i^{j'}-\delta_{i'}^j)\gamma^3D-\frac{ic}{2}\gamma^{3b}\gamma^3M_{3b}(\delta_i^{j'}-\delta_{i'}^j)-ic\gamma^3(U_i^{j'}+U_{i'}^j)\quad.}
The algebra will close only if
\eq{\delta_i^j=\delta_{i'}^{j'}\quad,}
since then we get the specific combinations of the generators $P_3-c^2K_3$, $P_a+c^2K_a$. 
Moreover, as discussed above $D$ cannot appear in the algebra, implying another constraint
\eq{\quad \delta_i^{j'}=\delta_{i'}^j\quad.}
This condition also ensures that $[U_i^{j'}+U_{i'}^j,\qq_{k k'}]$ is in the algebra.
With these constraints 
the commutator is simplified to
\eq{\{\qq_{i i'},\qq^{j j'}\}=-\onov{2}\delta_i^j\left[\gamma^3(P_3-c^2K_3)+\gamma^a(P_a+c^2K_a)  \right]-\frac{ic}{2}\gamma^{ab}\gamma^3M_{ab}\delta_i^{j'}-ic\gamma^3(U_i^{j'}+U_{i'}^j)\quad.}

Up to permutations, there are three possible solutions to the constraints, and we will analyze each of them separately :
\begin{table}[h]
\centering
\bgroup
\def\arraystretch{1.5}
\begin{tabular}{c c c c}
& \underline{  \textrm{I}  }  &\underline{   \textrm{II}  } &\underline{   \textrm{III}  }\\
$\qq_1$ = & $Q_1+ic\gamma^*\gamma^3S_1$ & $Q_1+ic\gamma^*\gamma^3S_1$ &  $Q_1+ic\gamma^*\gamma^3S_2$ \\
 $\qq_2$ = &  $Q_2+ic\gamma^*\gamma^3S_2$  & $Q_2+ic\gamma^*\gamma^3S_2$ & $Q_2+ic\gamma^*\gamma^3S_1$\\
$\qq_3$ =  & $Q_3+ic\gamma^*\gamma^3S_3$ & $Q_3+ic\gamma^*\gamma^3S_4$  &$Q_3+ic\gamma^*\gamma^3S_4$ \\
$\qq_4$ = & $ Q_4+ic\gamma^*\gamma^3S_4$ & $Q_4+ic\gamma^*\gamma^3S_3$ & $Q_4+ic\gamma^*\gamma^3S_3$ \\

\end{tabular}
\egroup
\end{table}

The naive analysis of the $R$-symmetry in the $2d$ superconformal algebra would be to check what subgroup of the entire $R$ group is consistent with the supercharges $\qq_{ii'}$. If we define the matrix $Z_i^{\ j}$ via $\qq_{i}=Q_i+ic\gamma^*\gamma^3Z_i^{\ j}S_j$, then the matrix $Z$ is invariant under $\mathcal{G}_R= SU(4),\ SU(3)\times U(1)$, and $\ SU(2)\times SU(2)\times U(1)$ for the cases $I,\ II,\ III$ respectively. The naive expectation is that the $2d$ $R$-symmetry will be a product of $\mathcal{G}_R$ with the $S^1$ isometry (or $S^q$ isometries in the general case). As we already saw in the $\mathcal{N}=2$ analysis, this is not the case. The mixture of the sphere isometries and the $R$-symmetry generators can modify the symmetry by central charges and $U(1)$ factors. In some cases we will see that not all of the generators appear on the right-hand side of anti-commutators of supercharges, and therefore will not be part of the $2d$ $R$-symmetry. For this reason, a more careful analysis needs to be done. 

\begin{enumerate}[I.]
\item \label{itm:eight}
This case is similar to the $\mathcal{N}=(0,4)$ case above, and the algebra takes the form
\eq{&\{\qq_i,\qq_j\}=\{\qq^i,\qq^j\}=0\quad,\\
&\{\qq_i,\qq^j\}=-\onov{2}\delta_i^j\left[\gamma^3(P_3-c^2K_3)+\gamma^a(P_a+c^2K_a)\right]-\frac{ic}{2}\gamma^{ab}\gamma^3M_{ab}\delta_i^{j}-2ic\gamma^3U_i^{j}\quad,\\
&[P_3-c^2K_3,\qq_i]=ic\qq_i\quad,\quad [P_3-c^2K_3,\qq^i]=-ic\qq^i\quad,\\
&[M_{ab},\qq_i]=-\onov{2}\gamma_{ab}\qq_i\quad,\quad [M_{ab},\qq^i]=-\onov{2}\gamma_{ab}\qq^i\quad,\\
&[P_a+c^2K_a,\qq_i]=ic\gamma^3\gamma_a\qq_i\quad,\quad[P_a+c^2K_a,\qq^i]=-ic\gamma^3\gamma_a\qq^i\quad,\\
&[U_i^j,\qq_k]=-\delta_k^j\qq_i+\onov{4}\delta_i^j\qq_k\quad,\quad
[U_i^j,\qq^k]=\delta_i^k\qq^j-\onov{4}\delta_i^j\qq^k\quad
.}
Notice that all unitary traceless R-symmetry generators $U_i^j$ appear in the algebra, and therefore the full $SU(4)$ remains as a two dimensional R-symmetry. The additional generator $P_3-c^2K_3$, which is the generator of the $U(1)$ symmetry on the circle $S^1$, also appears as an R-symmetry. This algebra is isomorphic to the $\mathcal{N}=(0,8)$ superconformal algebra, where the $\mathcal{N}=8$ algebra appearing is the one with a $U(4)$ R-symmetry.
\item
In this case it will be convenient to write the algebra in terms of generators $\qq_1, \qq_2$ and $\qq_{\pm}=\onov{\sqrt{2}}(\qq_3\pm\qq_4)$. We will use indices $i,j=1,2$, and write the $\pm$ explicitly. The fermionic commutation relations read
\eq{&\{\qq_i,\qq_j\}=\{\qq^i,\qq^j\}=0\quad,\\&\{\qq^\pm,\qq^\pm\}=\{\qq^\pm,\qq^\mp\}=\{\qq_\pm,\qq_\pm\}=\{\qq_\pm,\qq_\mp\}=\{\qq_\pm,\qq^\mp\}=0\quad,\\
&\{\qq_i,\qq^j\}=-\onov{2}\delta_i^j\left[\gamma^3(P_3-c^2K_3)+\gamma^a(P_a+c^2K_a)\right]-\frac{ic}{2}\gamma^{ab}\gamma^3M_{ab}\delta_i^{j}-2ic\gamma^3U_i^{j}\quad,\\
&\{\qq_+,\qq^+\}=
-\onov{2}\left[\gamma^3(P_3-c^2K_3)+\gamma^a(P_a+c^2K_a)\right]-\frac{ic}{2}\gamma^{ab}\gamma^3M_{ab}-ic\gamma^3(U_3^3+U_4^4+U_3^4+U_4^3)\quad,\\
&\{\qq_-,\qq^-\}=-\onov{2}\left[\gamma^3(P_3-c^2K_3)+\gamma^a(P_a+c^2K_a)\right]+\frac{ic}{2}\gamma^{ab}\gamma^3M_{ab}+ic\gamma^3(U_3^3+U_4^4-U_3^4-U_4^3)\quad,\\
&\{\qq_+,\qq^i\}=\sqrt{2}ic\gamma^3(U_3^i+U_4^i)\quad,\\
&\{\qq_i,\qq^+\}=\sqrt{2}ic\gamma^3(U_i^3+U_i^4)\quad,\\
&\{\qq_i,\qq^-\}=0\quad,\{\qq_-,\qq^i\}=0\quad.\\
}
The bosonic commutation relations of the Lorentz group with $\qq_i$ and $\qq^i$ are as in case \ref{itm:eight}, and with $\qq^\pm$ and $\qq_\pm$ as in \eqref{pm}.
The R-generators that appear in the commutation relations form a $SU(3)\times U(1)$ algebra, where $U_3^3+U_4^4-U_3^4-U_4^3$ is the generator of the $U(1)$ symmetry, and it commutes with the other eight generators.
This $U(1)$ generator is diagonal in the $\qq$'s with eigenvalues of $-\onov{2}$ for $\qq^+$ and $\qq^i$ and $\frac{3}{2}$ for $\qq^-$;  $\onov{2}$ for $\qq_+$ and $\qq_i$ and $-\frac{3}{2}$ for $\qq_-$. Together with the additional $U(1)$ generator $P_3-c^2K_3$ we can create a $U(1)$ generator $\onov{2}(P_3-c^2K_3)-ic(U_3^3+U_4^4-U_3^4-U_4^3)$ which commutes with $\qq^1,\qq^2,\qq^+,\qq_1,\qq_2,\qq_+$, but not with $\qq^-,\qq_-$. Another combination $\frac{3}{2}(P_3-c^2K_3)+ic(U_3^3+U_4^4-U_3^4-U_4^3)$ commutes with $\qq^-,\qq_-$, but not with the rest.
To summarize, we get an algebra in which the generators $\qq^1,\qq^2,\qq^+,\qq_1,\qq_2,\qq_+$, together with $U(3)$ R-symmetry generators, obey non-trivial commutation relations among themselves and (anti-)commute with $\qq^-,\qq_-$. The full algebra is isomorphic to the $\mathcal{N}=(2,6)$ superconformal algebra in two dimensions, with
R-symmetry group $U(1) \times U(3)$.
\item
In this case it will be convenient to use fermionic generators $\mathcal{Q}_{\pm}^{(1)}=\onov{\sqrt{2}}(\mathcal{Q}_1\pm\qq_2)$ and $\qq_{\pm}^{(2)}=\onov{\sqrt{2}}(\qq_3\pm\qq_4)$, for which we get the following algebra:
\eq{&\{\mathcal{Q}^{(1)}_+,\mathcal{Q}^{-(1)}\}=\{\mathcal{Q}^{(2)}_+,\mathcal{Q}^{-(2)}\}=\{\mathcal{Q}^{(1)}_+,\mathcal{Q}^{-(2)}\}=\{\mathcal{Q}^{(2)}_+,\mathcal{Q}^{-(1)}\}=0\quad,\\&\{\mathcal{Q}_+^{(1)},\qq^{+(1)}\}=-\onov{2}\gamma^3\Bigl(P_3-c^2K_3+2ic(U_1^2+U_2^1+U_1^1+U_2^2)\Bigr)-\onov{2}\gamma_aL^a\quad,\\&\{\mathcal{Q}_+^{(2)},\qq^{+(2)}\}=-\onov{2}\gamma^3\Bigl(P_3-c^2K_3+2ic(U_3^4+U_4^3+U_3^3+U_4^4)\Bigr)-\onov{2}\gamma_aL^a\quad,\\
	&\{\mathcal{Q}_+^{(1)},\qq^{+(2)}\}=-ic\gamma^3(U_1^4+U_1^3+U_2^3+U_2^4)\quad,\\
&\{\mathcal{Q}_+^{(2)},\qq^{+(1)}\}=-ic\gamma^3(U_4^1+U_3^1+U_3^2+U_4^2)\quad,\\
&\{\mathcal{Q}_-^{(1)},\qq^{-(1)}\}=-\onov{2}\gamma^3\Bigl(P_3-c^2K_3+2ic(U_1^2+U_2^1-U_1^1-U_2^2)\Bigr)-\onov{2}\gamma_a\bar{L}^a\quad,\\&\{\mathcal{Q}_-^{(2)},\qq^{-(2)}\}=-\onov{2}\gamma^3\Bigl(P_3-c^2K_3+2ic(U_3^4+U_4^3-U_3^3-U_4^4)\Bigr)-\onov{2}\gamma_a\bar{L}^a\quad,\\
&\{\mathcal{Q}_-^{(1)},\qq^{-(2)}\}=-ic\gamma^3(U_1^4-U_1^3+U_2^3-U_2^4)\quad,\\
&\{\mathcal{Q}_-^{(2)},\qq^{-(1)}\}=-ic\gamma^3(U_4^1-U_3^1+U_3^2-U_4^2)\quad,
	}
	with the bosonic commutation relations as in \eqref{pm}. We see that the algebra splits into two commuting sectors with individual $SU(2)$ R-symmetry generators. 
	There is also a central charge, $k_\theta=(P_3-c^2K_3)+ic(U_1^2+U_2^1+U_3^4+U_4^3)$, appearing on the right-hand side, which commutes with all the algebra and with the $SU(2)_+\times SU(2)_-$ R-symmetry. The supercharges $\qq^{\pm(1)}$ and $\qq^{\pm(2)}$ are doublets of $SU(2)_{\pm}$ respectively. This algebra is isomorphic to the two dimensional `small' $\mathcal{N}=(4,4)$ superconformal algebra, with an additional central charge $k_\theta$ (that cannot appear in the super-Virasoro extension of this algebra).

\end{enumerate}

\subsubsection{$\mathcal{N}=3$}

No superconformal field theories with $\mathcal{N}=3$ that do not have $\mathcal{N}=4$ are known, but we still include this algebraically consistent case for completeness. By straightforward generalizations of the previous cases, we can get here the $\mathcal{N}=(0,6)$, $\mathcal{N}=(2,4)$, $\mathcal{N}=(4,2)$ or $\mathcal{N}=(6,0)$ two dimensional superconformal algebras.

\subsection{Five dimensional theories on $AdS_4\times S^1$}
\label{d5s1}

As classified by Nahm \cite{Nahm:1977tg}, the only possible five dimensional superconformal algebra has $\mathcal{N}=1$ supersymmetry and  is called $F(4)$. We will use the real form $F^2(4)$ as in \cite{Bergshoeff:2001hc} to write the algebra:
\eql{F4}{&[M_{\mu\nu},Q_{i\al}]=-\onov{4}(\gamma_{\mu\nu}Q_i)_\al\quad,\quad [M_{\mu\nu},S_{i\al}]=-\onov{4}(\gamma_{\mu\nu}S_i)_\al\quad,\\
&[D,Q_{i\al}]=\onov{2}Q_{i\al}\quad,\quad [D,S_{i\al}]=-\onov{2}Q_{i\al}\quad,\\
&[K_\mu,Q_{i\al}]=i(\gamma_\mu S_i)_\al\quad,\quad [P_\mu,S_{i\al}]=-i(\gamma_\mu Q_{i})_\al\quad,\\
&\{Q_{i\al},Q_{j\be}\}=-\onov{2}\epsilon_{ij}(\gamma^\mu)_{\al\be}P_\mu\quad,\quad \{S_{i\al},S_{j\be}\}=-\onov{2}\epsilon_{ij}(\gamma^\mu)_{\al\be}K_\mu\quad,\\
&\{Q_{i\al},S_{j\be}\}=-\frac{i}{2}\left(\epsilon_{ij}C_{\al\be}D+\epsilon_{ij}(\gamma^{\mu\nu})_{\al\be}M_{\mu\nu}+3C_{\al\be}U_{ij}\right)\quad,\\
&[Q_{i\al},U_{kl}]=\epsilon_{i(k}Q_{l)\al}\quad,\quad [S_{i\al},U_{kl}]=\epsilon_{i(k}S_{l)\al}\quad,\quad [U_{ij},U^{kl}]=2\delta_{(i}^{(k}U_{j)}^{l)}\quad.}
In this algebra the supersymmetry generators $Q_{i\alpha}$ and the superconformal generators $S_{i\alpha}$ ($i=1,2$, $\alpha=1,2,3,4$) are symplectic Majorana spinors, with a total of eight real components.
The $U_{ij}$ generators form a $SU(2)$ R-symmetry algebra, and they are anti-hermitian and symmetric,
\eq{(U_i^j)^*=-U_j^i\quad,\quad U_{ij}=U_{ji}\quad.}
Unlike our conventions in four dimensions, here the indices $i,j=1,2$ are raised and lowered by $\epsilon_{ij}$ and $\epsilon^{ij}$ which satisfy
\eq{\epsilon_{12}=\epsilon^{12}=1\quad,\quad \epsilon_{jk}\epsilon^{ik}=\delta_j^i\quad.}
The charge conjugation matrix $C$, and also the matrices $C\gamma_a$, are anti-symmetric.

The algebra \eqref{F4} is quite similar to the $\mathcal{N}=2$ superconformal algebra in four dimensions, where we saw that the algebra closes on the isometries of  $AdS_{d-1}\times S^1$ for the following choices of supercharges:
\eq{\qq_i=Q_i+ic\gamma^{d-1}S_i}
or 
\eq{\qq_1=Q_1+ic\gamma^{d-1}S_2,\quad \qq_2=Q_2+ic\gamma^{d-1}S_1\quad.}
We may expect to have the same two options here, but it turns out that only one of them is consistent: the twisted choice with
\eq{\qq_1\equiv Q_1+ic\gamma_4S_2,\quad \qq_2\equiv Q_2+ic\gamma_4S_1\quad.}
As in the four dimensional case it will be convenient to work in the basis
\eq{\qq_{\pm}=\onov{\sqrt{2}}(\qq_1\pm\qq_2)\quad,}
in which the algebra takes the form
\eq{&\{\qq_+,\qq_+\}=\{\qq_-,\qq_-\}=0\quad,\\
&\{\qq_+,\qq_-\}=-c\gamma^4\gamma^{ab}M_{ab}+\onov{2}\gamma^a(P_a-c^2K_a)+\onov{2}\gamma^4\bigl(P_4+c^2K_4+3c(U_{22}-U_{11})\bigr)\quad.}
Our choice of supercharges breaks the $SU(2)$ R-symmetry to a $U(1)$ with the generator $U_{11}-U_{22}$. As in section \ref{d4N1}, one
linear combination of the isometry of $S^1$ and this unbroken $U(1)_R$ appears on the right-hand side algebra and acts as a 3d
$SO(2)_R$ generator:
\eq{&[P_4+c^2K_4+3c(U_{22}-U_{11}),\qq_{\pm}]=\pm 7c\qq_\pm\quad.} 
The other combination is a global symmetry that may or may not be broken.
The full algebra that we find is equivalent to the $\mathcal{N}=2$ three dimensional superconformal algebra $OSp(2|2,\mathbb{R})$ \citep{Park:1999cw}. 
Indeed, based on the amount of supersymmetry, this is the only possibility.

\subsection{Six dimensional theories on $AdS_5\times S^1$}

The largest space-time dimension consistent with superconformal symmetry is six \cite{Nahm:1977tg}. Assuming no higher spin conserved charges, there are two possibilities in this case, which are both chiral: the minimal $\mathcal{N}=(1,0)$ superconformal algebra, and the extended $\mathcal{N}=(2,0)$ superconformal algebra. Putting them on $AdS_5\times S^1$ should lead to a four dimensional superconformal algebra with half of the number of fermionic generators.

\subsubsection{$\mathcal{N}=(1,0)$}

The $6d$ algebra in this case is given in terms of symplectic Majorana-Weyl spinors by \cite{Coomans:2011ih}
\eq{&[M_{\mu\nu},Q_{i\al}]=-\onov{2}(\gamma_{\mu\nu}Q_i)_\al\quad,\quad [M_{\mu\nu},S_{i\al}]=-\onov{2}(\gamma_{\mu\nu}S_i)_\al\quad,\\
&[D,Q_{i\al}]=\onov{2}Q_{i\al}\quad,\quad [D,S_{i\al}]=-\onov{2}S_{i\al}\quad,\\
&[K_\mu,Q_{i\al}]=-(\gamma_\mu S_i)_\al\quad,\quad [P_\mu,S_{i\al}]=-(\gamma_\mu Q_{i})_\al\quad,\\
&\{Q_{i\al},Q^{j\be}\}=-\onov{2}\delta_{i}^{j}(\gamma^\mu)_{\al}^\be P_\mu\quad,\quad \{S_{i\al},S^{j\be}\}=-\onov{2}\delta_{i}^j(\gamma^\mu)_{\al}^\be K_\mu\quad,\\
&\{Q_{i\al},S^{j\be}\}=\frac{1}{2}\left(\delta_{i}^j\delta_\al^\be D+\delta_{i}^j(\gamma^{\mu\nu})_{\al}^\be M_{\mu\nu}+4\delta_{\al}^\be U_{i}^j\right)\quad,\\
&[U_i^j,Q^k]=\delta_i^kQ^j-\onov{2}\delta_i^jQ^k\quad,\quad [U_i^j,Q_k]=-\delta^j_kQ_i+\onov{2}\delta_i^jQ_k\quad,\\&[U_i^j,S^k]=\delta_i^kS^j-\onov{2}\delta_i^jS^k\quad,\quad [U_i^j,S_k]=-\delta^j_kS_i+\onov{2}\delta_i^jS_k\quad,\\&[U_i^j,U_k^l]=\delta_{i}^{l}U_{k}^{j}-\delta_k^jU_i^l\quad,}
where $i,j=1,2$, and $\alpha,\beta=1,\cdots,8$.
As in the five dimensional case, the spinor indices are raised and lowered by $\epsilon^{ij}$ and $\epsilon_{ij}$, respectively. The $U_i^j$'s are generators of the $SU(2)$ R-symmetry.

Again, this algebra is very similar in its structure to the $\mathcal{N}=2$ superconformal algebra in four dimensions. As in the five dimensional case of section \ref{d5s1}, only the twisted combinations $\qq_1=Q_1+ic\gamma^5S_2$ and $\qq_2=Q_2+ic\gamma^5S_1$ form a consistent algebra. The $R$-symmetry that is preserved is $U(1)$, and we also have the $U(1)$ isometry of $S^1$. The remaining subalgebra is isomorphic to the $\mathcal{N}=1$ superconformal algebra in four dimensions, $SU(2,2|1)$. One combination of $U(1)$'s $P_5-c^2K_5-4ic(U_2^{\ 1}+U_1^{\ 2})$ appears in this algebra as the $U(1)_R$ generator, and the other one may or may not be a global symmetry.

\subsubsection{$\mathcal{N}=(2,0)$}
\label{omega}
This algebra has an $USp(4) \simeq SO(5)$ R-symmetry, generated by the symmetric 2-form $U^{ij}$ ($i,j=1,2,3,4$). The indices are raised and lowered by the antisymmetric invariant tensors $\Omega_{ij}$ and $\Omega^{ij}$, which satisfy $\Omega^{ij}\Omega_{jk}=\delta^i_k$. In order to perform computations, we choose a specific representation for these matrices: 
\eq{\Omega^{41}=\Omega^{23}=-\Omega^{14}=-\Omega^{32}=1.} 
The algebra then takes the form
\eq{&[M_{\mu\nu},Q^i_{\al}]=-\onov{4}(\gamma_{\mu\nu}Q^i)_\al,\quad [M_{\mu\nu},S^i_{\al}]=-\onov{4}(\gamma_{\mu\nu}S^i)_\al\quad,\\
&[D,Q^i_{\al}]=\onov{2}Q^i_{\al},\quad [D,S^i_{\al}]=-\onov{2}S^i_{\al}\quad,\\
&[K_\mu,Q^i_{\al}]=(\gamma_\mu S^i)_\al,\quad [P_\mu,S^i_{\al}]=(\gamma_\mu Q^{i})_\al\quad,\\
&\{Q^i_{\al},Q^j_{\be}\}=-2(\gamma^\mu)_{\al\be}\Omega^{ij} P_\mu,\quad \{S^i_{\al},S^j_{\be}\}=-2(\gamma^\mu)_{\al\be}\Omega^{ij} K_\mu\quad,\\
&\{Q^i_{\al},S^j_{\be}\}=-2C_{\al\be}\left(\Omega^{ij}D+4U^{ij}\right)-2(\gamma_{\mu\nu})_{\al\be}\Omega^{ij}M^{\mu\nu}\quad,\\
&[U_{ij},Q_k]=-\Omega_{k(i}Q_{j)},\quad [U_{ij},S_k]=-\Omega_{k(i}S_{j)}\quad,\\&[U_{ij},U_{kl}]=\Omega_{i(k}U_{l)j}+\Omega_{j(k}U_{l)i}\quad.}

Naively there are three choices of combining the supercharges to get a superconformal algebra on $AdS_5\times S^1$, just as in the $\mathcal{N}=4$ case in four dimensions. But it turns out that only for two of them the algebra closes; these are
\eql{qq6}{&\qq_1=Q_1+ic\gamma^5S_2,\quad \qq_2=Q_2+ic\gamma^5S_1,\quad \qq_3=Q_3+ic\gamma^5S_4,\quad \qq_4=Q_4+ic\gamma^5S_3\quad,\\
&\qq^1=Q^1-ic\gamma^5S^2,\quad \qq^2=Q^2-ic\gamma^5S^1,\quad \qq^3=Q^3-ic\gamma^5S^4,\quad \qq^4=Q^4-ic\gamma^5S^3\quad,}
and
\eq{&\qq_1=Q_1+ic\gamma^5S_4,\quad \qq_4=Q_4+ic\gamma^5S_1,\quad \qq_3=Q_3+ic\gamma^5S_2,\quad \qq_2=Q_2+ic\gamma^5S_3\quad,\\
&\qq^1=Q^1-ic\gamma^5S^4,\quad \qq^4=Q^4-ic\gamma^5S^1,\quad \qq^3=Q^3-ic\gamma^5S^2,\quad \qq^2=Q^2-ic\gamma^5S^3\quad.}
These two options turn out to give the same algebra, in which the $6d$ R-symmetry breaks to a $U(2)$ symmetry with
\eq{U_+=U_{44}-U_{33},\quad U_-=U_{22}-U_{11},\quad U_z=U_{14}-U_{32},\quad T=U_{24}-U_{13}\quad.}
$U_+,U_-$ and $U_z$ satisfy the $SU(2)$ algebra, and $T$ commutes with them. Defining combinations of the $\qq$'s of \eqref{qq6} as in case I of section \ref{d4N4}, they act on the supercharges in the following way
\eq{&[U_z,\qq_{\pm}^i]=(\sigma_z)^{i}_j\qq_{\pm}^j,\quad[U_+,\qq_{\pm}^1]=\qq_{\pm}^2,\quad[U_-,\qq_{\pm}^2]=\qq_{\pm}^1,\quad[T,\qq_{\pm}^i]=\pm\qq_{\pm}^i\quad.}
We also have the extra $U(1)$ generator $\dth=P_5-c^2 K_5$. One combination of this with $T$ acts as the $U(1)$ R-generator in the four dimensional superconformal algebra, and the other may or may not be a $U(1)$ global symmetry.
The algebra we find is equivalent to the four dimensional $\mathcal{N}=2$ superconformal algebra $SU(2,2|2)$; again this is the only possibility based on the counting of supercharges. Theories realizing this construction were discussed in \cite{Aharony:2015zea}.
  
  \subsection{Field theories on $AdS_{d-2}\times S^2$}

in this case we wish our symmetries to commute with the $SO(3)$ isometry group of the $S^2$ factor. Using similar manipulations to the
$AdS_{d-1}\times S^1$ cases, we can single out the last two space-time dimensions by using two gamma matrices in the form of the conserved supercharges on $AdS_{d-2}\times S^2$, $\qq_i = Q_i+ic\gamma^{(d-2)(d-1)}S_{i'}$. This ansatz turns out to give a superconformal algebra that preserves the isometries of $AdS_{d-2}\times S^2$.
Since we are not studying $AdS_2$ here, we can choose $d=5$ or $d=6$; however $d=6$ turns out to be impossible
(see Appendix \ref{6d}) so we only have one case.

    \subsubsection{Five dimensional field theories on $AdS_3\times S^2$}

 As in \eqref{F4}, we consider the $F(4)$ superconformal algebra. This time we choose as conserved supercharges $\qq_i=Q_i+ic\gamma^{34}S_{i}$, which turns out to be the only consistent choice giving a closed subalgebra. We will denote the coordinates $\mu=3,4$ by $A,B,\cdots$ and $\mu=0,1,2$ by $a,b,\cdots$. The fermionic part of the algebra is

\eql{ads3s2}{&\{\qq_1,\qq^2\}=-3c\gamma^{34}\tensor{U}{_{1}^2},\quad \{\qq_2,\qq^1\}=-3c\gamma^{34}\tensor{U}{_{2}^1}\\
&\{\qq_1,\qq^1\}=\onov{2}\gamma^A\left(P_A+c^2K_A\right)+\onov{2}\gamma^a\left(P_a-c^2K_a\right)-cM_{ab}\gamma^{ab}\gamma^{34}+2cM_{34}+3c\gamma^{34}\tensor{U}{_{1}^1}\\
&\{\qq_2,\qq^2\}=\onov{2}\gamma^A\left(P_A+c^2K_A\right)+\onov{2}\gamma^a\left(P_a-c^2K_a\right)-cM_{ab}\gamma^{ab}\gamma^{34}+2cM_{34}-3c\gamma^{34}\tensor{U}{_{2}^2}.}
The $SO(2,2)=SL(2,R)\times SL(2,R)$ isometry group of $AdS_3$ is now generated by 
\eq{&\gamma^a(P_a-c^2K_a)-cM_{ab}\gamma^{ab}\gamma^{34}=\gamma^a L_a\quad,\\
&\gamma^a(P_a-c^2K_a)+cM_{ab}\gamma^{ab}\gamma^{34}=\gamma^a \bar L_a\quad.}
Only the $L_a$ appear on the right-hand side of \eqref{ads3s2}, therefore the two dimensional superconformal algebra is chiral.
$P_A+c^2K_A$ and $M_{34}$ generate the $SO(3)$ isometry group of the sphere, and $U_1^2$, $U_2^1$ and $U_1^1=-U_2^2$ generate an $SU(2)_R$ symmetry.
In the two dimensional superconformal algebra, the sphere generators join with the $R$-symmetry generators to form an $SU(2)\times SU(2)$ R-symmetry. This algebra turns out to be the `large' $\mathcal{N}=(0,4)$ superconformal algebra~\citep{Ali:1999ut} \footnote{The R-symmetry of the `large' $\mathcal{N}=4$ superconformal algebra is $SU(2)\times SU(2)\times U(1)$, but only $SU(2)\times SU(2)$ appears in the global part of the superconformal algebra (as opposed to the full Virasoro).}. 

\subsection{Field theories on $AdS_{d-3}\times S^3$}

Here the only examples are the six dimensional ones. We use the same conventions for the $6d$ superconformal algebras as above. Following the previous  sections we propose the conserved supercharges to be of the form $Q_i+ic\gamma^{(d-3)(d-2)(d-1)}S_{i'}$, and we denote the indices $\mu=3,4,5$ by $A,B,\cdots$, and the indices $\mu=0,1,2$ by $a,b,\cdots$.

\subsubsection{Six dimensional $\mathcal{N}=(1,0)$ theories}

The only consistent choice for combining the supercharges is the diagonal choice,
\eq{\qq_1=Q_1+ic\gamma^{345}S_1,\quad \qq_2=Q_2+ic\gamma^{345}S_2.}
These obey the algebra
\eq{\{\qq_1,\qq^1\}=&-\onov{2}\gamma^a(P_a-c^2K_a)-\onov{2}\gamma^A(P_A+c^2K_A)+\frac{ic}{2}\gamma^{345}\gamma^{ab}M_{ab}\\&+\frac{ic}{2}\gamma^{345}\gamma^{AB}M_{AB}+4ic\gamma^{345}U_1^{\ 1}\quad,\\
\{\qq_2,\qq^2\}=&-\onov{2}\gamma^a(P_a-c^2K_a)-\onov{2}\gamma^A(P_A+c^2K_A)+\frac{ic}{2}\gamma^{345}\gamma^{ab}M_{ab}\\&+\frac{ic}{2}\gamma^{345}\gamma^{AB}M_{AB}+4ic\gamma^{345}U_2^{\ 2}\quad,\\
\{\qq_1,\qq^2\}=& 4ic\gamma^{345}U_1^{\ 2},\quad \{\qq_2,\qq^1\}=4ic\gamma^{345}U_2^{\ 1}\quad.}
The generators $(P_A+c^2K_A)+ic\epsilon_{ABC}M^{BC}$ and $U_i^{\ j}$ form an $SU(2)\times SU(2)$  R-symmetry in the two dimensional superconformal algebra. As in the previous case, the full algebra turns out to be
the `large' $\mathcal{N}=(0,4)$ superconformal algebra in two dimensions.
The other three $S^3$ rotation generators $P_A+c^2K_A-ic\epsilon_{ABC}M^{AB}$ commute with the supercharges and may or may not be a global symmetry.

\subsubsection{Six dimensional $\mathcal{N}=(2,0)$ theories}

In this case there are two options to form a consistent algebra. The first case is the diagonal case 
\eq{\qq_i=Q_i+ic\gamma^{345}S_{i},\quad\qq^i=Q^i-ic\gamma^{345}S^{i}.}
The R-symmetry in this case  consists of an $SU(2)$ subgroup of the sphere $SO(4)$ isometries generated by $P_A+c^2K_A+ic\epsilon_{ABC}M^{BC}$, and of the full $6d$ $R$-symmetry $USp(4)\sim SO(5)$, and we have 8 chiral supercharges. We obtain a $\mathcal{N}=(0,8)$ superconformal algebra that is different from the one we encountered before; this algebra is classified as case (III) in 
\citep{Fradkin:1992bz}. The other three $SO(4)$ generators again commute with the supercharges and may or may not be a global symmetry.

The second  option is to split the generators into two pairs. For the representation of $\Omega$ that we chose in section \ref{omega} they are
\eq{&\qq_1=Q_1+ic\gamma^{345}S_{3},\quad \qq_2=Q_2+ic\gamma^{345}S_{4},\\
&\qq_3=Q_3+ic\gamma^{345}S_{1},\quad \qq_4=Q_4+ic\gamma^{345}S_{2}.}
Here the supercharges close on the entire $SO(4)$ sphere isometries, and our choice preserves an $SO(4)$ subgroup of the $USp(4)\simeq SO(5)$ $R$-symmetry. Altogether we obtain the `large' $\mathcal{N}=(4,4)$ superconformal algebra with $SO_L(4)\times SO_R(4)$ $R$-symmetry.
Each one of the $R$-symmetry groups $SO(4)_{L/R}$ acts only on left/right-handed supercharges, and is generated by three out of the $S^3$ isometries and three out of the preserved six dimensional $R$-symmetry generators. 

\subsection{Field theories on $AdS_d$}

This is the final possibility, which is related by the conformal transformation discussed above to flat space with a codimension one boundary.
In this case our ansatz for the conserved supercharges is simply $\qq=Q+ic\gamma^*S$. The supersymmetry algebra should be a $(d-1)$-dimensional superconformal algebra. For the $d=6$ $\mathcal{N}=(2,0)$ case it is clear just by counting supercharges that this is not possible, and in fact it is easy to see (essentially by chirality arguments) that one cannot preserve half of the supersymmetry for any $6d$ theory with chiral supersymmetry on $AdS_6$ (see Appendix \ref{6d}). So, we will analyze the three, four and five dimensional cases.

\subsubsection{Four dimensional $\mathcal{N}=1$}

Four dimensional supersymmetric theories on $AdS_4$ were discussed in the past, for instance in \cite{Zumino:1977av,Ivanov:1979ft,Ivanov:1980vb,Sakai:1984nc,Burgess:1984rz,Burgess:1984ti,Burges:1985qq}, and also more recently in \cite{Adams:2011vw,Aharony:2010ay,Aharony:2012jf}.
For the ${\mathcal{N}}=1$ case the algebra we obtain must be the three dimensional $\mathcal{N}=1$ superconformal algebra, with no $R$-symmetry,
and indeed we find
\eql{4dads4}{&\{\qq,\qq\}=-\onov{2}\gamma^\mu(P_\mu+c^2K_\mu)-\frac{ic}{2}\gamma^{\mu\nu}\gamma^*M_{\mu\nu},\\
&[P_\mu+c^2K_\mu,\qq]=ic\gamma^*\gamma_\mu\qq,\\
&[M_{\mu\nu},\qq]=-\onov{2}\gamma_{\mu\nu}\qq.}
The extra $U(1)_R$ symmetry that we have in four dimensions is broken by the choice of the combination of supercharges that appears in \eqref{4dads4}; this must happen because a codimension one boundary reflects left-handed fermions into right-handed ones.

\subsubsection{Four dimensional $\mathcal{N}=2$}

As we saw in previous similar cases, also here there are two options, the diagonal case $\qq_i=Q_i+ic\gamma^*S^i$ and the twisted case $\qq_1=Q_1+ic\gamma^*S^2,\quad \qq_2=Q_2+ic\gamma^*S^1$. As can be seen by a change of basis, the two options turn out to be equivalent and obey the algebra 
\eq{&\{\qq_1,\qq^1\}=\{\qq_2,\qq^2\}=-\onov{2}\gamma^\mu(P_\mu+c^2K_\mu),\quad \{\qq_1,\qq^2\}=\{\qq_2,\qq^1\}=0\quad,\\
&\{\qq_1,\qq_2\}=ic(U_1^2-U_2^1),\quad \{\qq_1,\qq_1\}=\{\qq_2,\qq_2\}=\frac{ic}{2}\gamma^{\mu\nu}M_{\mu\nu}\quad,\\
&[U_1^2-U_2^1,\qq_1]=\qq_2,\quad [U_1^2-U_2^1,\qq_2]=-\qq_1\quad.}
This algebra is isomorphic to the three dimensional $\mathcal{N}=2$ superconformal algebra with an $SO(2)$ R-symmetry, with $U_1^2-U_2^1$ as its generator; the other generators of the $4d$ $U(2)$ R-symmetry are broken.

\subsubsection{Four dimensional $\mathcal{N}=4$}
Also in this case all three options of combining the supercharges $Q_i$ with the superconformal charges $S^{i'}$ turn out to give the same algebra 
\eq{&\{\qq_i,\qq^j\}=-\onov{2}\delta_i^j\gamma^\mu(P_\mu+c^2K_\mu),\quad \{\qq_i,\qq_j\}=\frac{ic}{2}\delta_{ij}\gamma^{\mu\nu}M_{\mu\nu}+ic(U_i^j-U_j^i),\\
&[U_i^j-U_j^i,\qq_k]=\delta_k^iQ_j-\delta_k^jQ_i.}
This is equivalent to the three dimensional $\mathcal{N}=4$ superconformal algebra with $SO(4)$ $R$-symmetry. Again the $4d$ $SU(4)$
R-symmetry is broken to $SO(4)$. Similarly, the $4d$ $\mathcal{N}=3$ case leads to a $3d$ $\mathcal{N}=3$ superconformal algebra.

\subsubsection{Five dimensional theories on $AdS_5$}

In the five dimensional case only one out of the two natural options is consistent. This is the twisted choice, $\qq_1=Q_1+icS_2$ and $\qq_2=Q_2+icS_1$. The algebra is then the $\mathcal{N}=1$ four dimensional superconformal algebra, with a $U(1)$ R-symmetry generated by $U_{11}-U_{22}$.

\subsubsection{Three dimensional theories on $AdS_3$}

The three dimensional superconformal algebra is
\eq{&\{Q^{i\al},\bar{Q}_{j\be}\}=2\delta^i_{\ j}\gamma^{\mu\al}_{\ \be}P_\mu\ ,\ \{S^{i\al},\bar{S}_{j\be}\}=2\delta^i_{\ j}\gamma^{\mu\al}_{\ \be}K_\mu,\\
&[M_{\mu\nu},Q^i]=\frac{i}{2}\gamma_{\mu\nu}Q^i\ ,\ [M_{\mu\nu},S^i]=\frac{i}{2}\gamma_{\mu\nu}S^i\ ,\ [P_\mu,S^i]=-\gamma_\mu Q^i\ ,\ [K_\mu,Q^i]=-\gamma_\mu S^i,\\
&\{Q^{i\al},\bar{S}_{j\be}\}=-i\delta^i_{\ j}\left(2\delta^\al_{\ \be}D+(\gamma^{\mu\nu})^\al_{\ \be}M_{\mu\nu}\right)+2i\delta^\al_{\ \be}A^i_{\ j},\\
&[A_{ij},A_{kl}]=i(\delta_{ik}A_{jl}-\delta_{il}A_{jk}-\delta_{jk}A_{il}+\delta_{jl}A_{ik})\ ,\ [A_{ij},Q^k]=i(\delta_i^{\ k}\delta_{jl}-\delta_j^{\ k}\delta_{il})Q^l.}
Here $A_{ij}$ are $SO({\cal N})$ generators, $i,j=1,\cdots,{\cal N}$. For more conventions see \citep{Park:1999cw}.

As in previous cases, in order to close the algebra on the $AdS_3$ isometries, we define the supercharge $\qq^a=Q^{a}+cS^{a'}$ that gives the commutator
\eq{&\{\qq^{a\al},\bar{\qq}_{b\be}\}=\{Q^{a\al},\bar{Q}_{b\be}\}+c\{S^{a'\al},\bar{Q}_{b\be}\}+c\{Q^{a\al},\bar{S}_{b'\be}\}+c^2\{S^{a'\al},\bar{S}_{b'\be}\}\\
&=2\gamma^{\mu \al}_{\ \ \be}(\delta^a_{\ b}P_\mu+c^2\delta^{a'}_{\ b'}K_\mu)-2ic(\delta^a_{\ b'}-\delta^{a'}_{\ b})\delta^\al_{\ \be}D-ic(\delta^a_{\ b'}+\delta^{a'}_{\ b})(\gamma^{\mu\nu})^\al_{\ \be}M_{\mu\nu}+2ic\delta^\al_{\ \be}(A^a_{\ b'}+A^{a'}_{\ b}).}
For some ${\cal N}$, we can choose $n$ diagonal and $2m$ twisted supercharges where $n+2m={\cal N}$, such that we define 
\eq{&\qq^i=Q^i+cS^i\ ,\ i=1,\cdots,n\\
&\qq^a=Q^a+cS^{a+m}\ ,\ a=n+1,\cdots, n+m\\
&\qq^{a'}=Q^{a+m}+cS^{a}\ ,\ a=n+1,\cdots,n+m.}
The algebra is then
\eq{&\{\qq^i,\qq_i\}=2\gamma^\mu(P_\mu+c^2K_\mu)-2ic\gamma^{\mu\nu}M_{\mu\nu},\\
&\{\qq^i,\qq_{j\neq i}\}=4icA^i_{\ j},\\
&\{\qq^a,\qq_a\}=2\gamma^\mu(P_\mu+c^2K_\mu),\\
&\{\qq^a,\qq_{b\neq a}\}=2ic(A^a_{\ b+m}+A^{a+m}_{\ b}),\\
&\{\qq^a,\qq_{a'}\}=-2ic\gamma^{\mu\nu}M_{\mu\nu},\\
&\{\qq^a,\qq_{b'\neq a'}\}=2ic(A^a_{\ b}+A^{a+m}_{\ b+m}),\\
&\{\qq^i,\qq_a\}=\{\qq^i,\qq_{a'}\}=2ic(A^i_{\ a}+A^i_{\ a+m}).}
From chirality analysis, we get the $\mathcal{N}=(m,n+m)$ $2d$ superconformal algebra. The $n+m$ right handed spinors are $\qq^i, \onov{\sqrt{2}}(\qq^a+\qq^{a'})$, and the $m$ left-handed spinors are $\onov{\sqrt{2}}(\qq^a-\qq^{a'})$.
 There are $\frac{n(n-1)}{2}$ $A^i_{\ j}$ generators, $\frac{m(m-1)}{2}$ $(A^a_{\ b+m}+A^{a+m}_{\ b})$ generators, $\frac{m^2-m}{2}$ $(A^a_{\ b}+A^{a+m}_{\ b+m})$ generators and $nm$ $(A^i_{\ a}+A^i_{\ a+m})$ generators that are not broken, so altogether we have $\frac{(n+m)(n+m-1)}{2}+\frac{m(m-1)}{2}$ generators, that give an $SO(n+m)\times SO(m)$ $R$-symmetry algebra which acts on the left-handed and right-handed spinors.

The $\mathcal{N}=(0,n+m)$ subgroup involves the $n+m$ right handed spinors, the three $P_\mu+c^2K_\mu-ic\epsilon_{\mu\nu\rho}M^{\nu\rho}$ $SL(2)_R$ isometries, and an $SO(n+m)$ R symmetry made out of $A^i_{\ j}, (A^i_{\ a}+A^i_{\ a+m})$, and $(A^a_{\ b+m}+A^{a+m}_{\ b}+A^a_{\ b}+A^{a+m}_{\ b+m})$.
The $\mathcal{N}=(m,0)$ subgroup involve the $m$ left handed spinors, the three $P_\mu+c^2K_\mu+ic\epsilon_{\mu\nu\rho}M^{\nu\rho}$
 $SL(2)_L$ isometries, and an $SO(m)$ $R$-symmetry made out of $(A^a_{\ b+m}+A^{a+m}_{\ b}-A^a_{\ b}-A^{a+m}_{\ b+m})$.
The two subgroups (anti-)commute with each other.

\section{Supersymmetric field theory on $AdS_3\times S^1$}
\label{AdS3S1}

In this section, we present some specific examples illustrating the results of section \ref{SCalgebra} for the cases of $4d$ $\mathcal{N}=1,2$ supersymmetric field theories on $AdS_3\times S^1$. As we showed there, the different possibilities of putting  $\mathcal{N}=1,2$ SUSY on $AdS_3\times S^1$ give supersymmetry algebras equivalent to the $\mathcal{N}=(0,2),(2,0),(0,4),(4,0)$, and $(2,2)$ two dimensional superconformal algebras. 

We will show explicitly how to write the actions and transformation rules for different $4d$ multiplets.
For $\mathcal{N}=1$ we use the simple notations of new minimal supergravity (SUGRA) and the results of \cite{Dumitrescu:2012ha}. For $\mathcal{N}=2$, we build the actions and transformation rules using the superconformal approach discussed in section \ref{SCalgebra}, starting from an $\mathcal{N}=2$ superconformal theory and coupling it to superconformal gravity. The different choices for the supercharges $\qq_i$ correspond to relations $\eta(\zeta)$, where $\zeta,\eta$ are parameters related to the $Q$ and $S$ transformations, respectively. By starting from superconformal field theories on flat space and plugging in the relations $\eta(\zeta)$, we get the correct Killing spinor equations, action, and transformation rules (see appendix \ref{KSE} for more details).

We focus on free theories for which we will explicitly construct the  action and boundary conditions on $AdS_3\times S^1$ that preserve all of the supercharges, and study the spectrum of the $2d$ superconformal algebras. Unlike in the previous section, here we allow for different radii for $AdS_3$ and $S^1$, in order to show that one can still preserve the same supersymmetry algebras also in this case. We also allow for $4d$ field theories that are not necessarily conformal, though most of our examples will be conformal.

\subsection{Four dimensional $\mathcal{N}=1$ theories on $AdS_3\times S^1$}
\label{algebra}

In the previous section we analyzed the supersymmetry of  $\mathcal{N}=1$ theories on $AdS_3\times S^1$ algebraically. Another general way to study such theories is to couple them to background fields of
 new minimal supergravity, and to use the results of \citep{Dumitrescu:2012ha}. We will show explicitly that the two consistent values for the background supergravity fields result in $2d$ $\mathcal{N}=(0,2)$ and $\mathcal{N}=(2,0)$ superconformal algebras. We use the metric 
\eq{ds^2=\frac{L^2}{r^2}(-dt^2+dx^2+dr^2)+R^2d\te^2,}
where $\te$ is the coordinate on $S^1$ with $\te\sim\te+2\pi$, $R$ and $L$ are the radii of the $S^1$ and $AdS_3$ respectively, and for $r\rightarrow 0$ we reach the boundary of AdS.
The curved space sigma matrices are related to the flat ones by
\eq{\sigma_{t,x,r}=\frac{L}{r}\sigma_{0,1,2},\quad \sigma_\te=R\sigma_3.}
For spinors and sigma matrices conventions, we follow \cite{Wess:1992cp}. 

The classification of geometries preserving different numbers of supercharges for four dimensional $\mathcal{N}=1$ theories on various manifolds can be found in  \citep{Dumitrescu:2012ha, Festuccia:2011ws}.
 Following their work we couple the theory to the new minimal supergravity multiplet \citep{Sohnius:1981tp} which contains, in addition to the physical graviton $\gmn$ and gravitino $\Psi_\mu^\al$, the following auxiliary fields: a $U(1)_R$ gauge field $A_\mu$ and the 1-form $V^\mu=\onov{4}\epsilon^{\mu\nu\rho\lambda}\dn B_{\rho\lambda}$.
 The conditions for preserving all four supercharges are given by
\begin{align}
\label{Vcond}&\nabla_\mu V_\nu=0,\nonumber\\
&\partial_{[\mu}A_{\nu]}=0,\nonumber\\
&W_{\mu\nu\kappa\lambda}=0,\nonumber\\
&\mathcal{R}_{\mu\nu}=-2(V_\mu V_\nu-g_{\mu\nu} V_\rho V^\rho),
\end{align}
where $W_{\mu\nu\kappa\lambda},\mathcal{R}_{\mu\nu}$ are the Weyl and Ricci tensors of the metric $g_{\mu \nu}$, respectively.
When these conditions are satisfied, there are four independent solutions to the Killing spinor equations 
\begin{align}
\label{killing}
(\nabla_\mu-iA_\mu)\zeta&=-iV_\mu\zeta-iV^\nu\sigma_{\mu\nu}\zeta,\nonumber\\
(\nabla_\mu+iA_\mu)\zeb&=iV_\mu\zeb+iV^\nu\bar\sigma_{\mu\nu}\zeb\quad .
\end{align}
The superalgebra then will be
\eql{susyalg}{\{\delta_\zeta,\delta_{\zeb}\}=2i\delta_K,\quad
\{\delta_\zeta,\delta_\zeta\}=\{\delta_{\zeb},\delta_{\zeb}\}=0,}
where $\delta_K$ is the $R$-covariant Lie derivative along the Killing vector $K^\mu=\zeta\sigma^\mu\zeb$,
\eq{\delta_K=\mathfrak{L}_K-i\hat{q}K^\mu A_\mu,}
and $\hat{q}$ is the generator of the $U(1)_R$ symmetry.

We begin by determining the values of the background fields  $V_\mu$ and $A_\mu$. From \eqref{Vcond} we find 
\eq{V^\mu V_\mu=\onov{6}\mathcal{R}=\onov{L^2}.}
In order to preserve the isometries of our spacetime, $V^\mu$ and $A^\mu$ must take values in the $S^1$ direction. We then get two solutions
\eq{V_\mu=\pm\frac{R}{L}\delta_\mu^\te.}
From the Killing spinor equations and the requirement that the spinors should be single-valued (see appendix \ref{Rgauge}), we find that the allowed values for $A_\mu$ are
\eq{ A_\mu=\left(1+\frac{nL}{R}\right)V_\mu,\quad n\in\mathbb{Z}.}
The parameter $n$ here corresponds to a large gauge transformation of the background $U(1)_R$ field around the circle, which is essentially the same as shifting the momentum generator around the circle (normalized to be an integer) by $n$ times the R-charge. The effect of this is discussed in appendix \ref{Rgauge}. It has no effect on the supersymmetry algebra, so from here on we will set $n=0$.

Note that this construction only works when we have a $U(1)_R$ symmetry; for superconformal theories this is guaranteed, but for other theories it is a necessary condition for preserving all supercharges on $AdS_3\times S^1$.

We can now solve \eqref{killing} to get an explicit form for the Killing spinors $\zeta$ and $\zeb$:

\begin{tabular}{ll}
 \\[0.1pt]
\multirow{2}{*}{$V_\mu=-\frac{R}{L}\delta_\mu^\theta$}&$\qquad\zeta=\zeta_R=ar^{\onov{2}} \mat{1\\1}+(b+iaz)r^{-\onov{2}}\mat{1\\-1}$\\
&$\qquad\zeb=\zeb_R=-\bar{a}r^{\onov{2}}\mat{1&1}+(\bar{b}+i\bar{a}z)r^{-\onov{2}}\mat{1&-1}$\\
 \\[0.1pt]
\hline
 \\[0.1pt]
\multirow{2}{*}{$V_\mu=\frac{R}{L}\delta_\mu^\theta$}&$\qquad\zeta=\zeta_L=ar^{\onov{2}}\mat{1\\-1}+(b+ia\bar z)r^{-\onov{2}}\mat{1\\1}$\\
&$\qquad\zeb=\zeb_L=\bar{a}r^{\onov{2}}\mat{1&-1}-(\bar{b}+i\bar{a}\bar z)r^{-\onov{2}}\mat{1&1}$\\
 \\[0.1pt]
\end{tabular}\\

Here $a,\bar{a},b$ and $\bar{b}$ are Grassmanian parametrizations of the components of the Killing spinors which correspond to the four independent supercharges, $z\equiv x+t$ and $\zb\equiv x-t$ are coordinates in the spatial directions of the boundary of $AdS_3$, and the subscripts $L,R$ denote left/right-handed solutions.

From now on we will focus on the right-handed solution $\zeta_R,\zeb_R$; a similar analysis can be done for the left-handed one.
Using the explicit form of the spinors, we can compute the Lie derivative $\delta_K$ acting on different fields.

For example, when acting on a scalar (which can have some non-zero R-charge as the eigenvalue of $\hat q$), the Lie derivative takes the simple form $\lag_K=K^\mu\dm$, resulting in the following commutators of the generators:
 \eql{salgebra}{\{\delta_{\bar{b}},\delta_b\}&=\frac{4i}{L}(\dx-\dt),\\
\{\delta_{\bar{a}},\delta_a\}&=-\frac{4iz^2}{L}(\dx-\dt)-\frac{8iz}{L}(r\dr)+O(r^2),\\
  \{\delta_{\bar{a}},\delta_b\}&=-\frac{4z}{L}(\dx-\dt)-\frac{4}{L}(r\dr)+\frac{4i}{R}\left(\partial_\te+\frac{i\hat{q}R}{L}\right),\\
   \{\delta_{\bar{b}},\delta_a\}&=-\frac{4z}{L}(\dx-\dt)-\frac{4}{L}(r\dr)-\frac{4i}{R}\left(\partial_\te+\frac{i\hat{q}R}{L}\right).}
Taking $r\rightarrow 0$ we can identify this with the two dimensional $\mathcal{N}=(0,2)$ superconformal algebra, as in section \ref{d4N1}, with generators
\eql{identification}{&\Delta=r\dr ,\quad \frac{3}{2}\hat{R}=\frac{L}{R}\left(\dth+\frac{i\hat{q}R}{L}\right),\\
&Q=\sqrt{\frac{L}{2}}\delta_b\ ,\quad \bar Q=\sqrt{\frac{L}{2}}\delta_{\bar{b}},\\
&S=-i\sqrt{\frac{L}{2}}\delta_{\bar{a}}\ ,\quad \bar{S}=i\sqrt{\frac{L}{2}}\delta_a.}
Here  $\Delta=h_L+h_R$ is the sum of the left and right dimensions (which are equal for scalars), and $\hat{R}$ is the $U(1)_R$ generator of the two dimensional superconformal algebra. We can repeat the procedure for higher spin fields and find also the spin, $s=h_R-h_L$. 
We find that for a bulk Weyl spinor with $\psi_\pm=\mat{1\\\pm1}$, the $2d$ spin value is $\mp\onov{2}$, and for a vector $v_x\pm v_t$, the $2d$ spin value is $\pm1$. 

Thus, we find also in the explicit field theory language the same algebraic structure as in the previous section.
Note in particular that, as in section \ref{d4N1}, the $2d$ $R$-charge $\hat R$ is a linear combination of the KK momentum and the $4d$ $R$-charge; the
specific combination we had in section \ref{d4N1} arises here for $L=R$ (which we assumed in the previous section).
If we choose the opposite sign for $V_\mu$, we similarly get the $\mathcal{N}=(2,0)$ superconformal algebra.

\subsection{Free field theory on $AdS_3\times S^1$}\label{N0}

In this section we analyze the spectrum and boundary conditions of different fields on $AdS_3\times S^1$. For related work see, for example, \citep{marolf2006boundary} and references within. We begin with free fields, and later join them into supersymmetry multiplets.

\subsubsection{Scalar}
The bulk action for a free massless scalar of R-charge $q$ (which is the bottom component of a chiral multiplet) coupled to the new minimal supergravity auxiliary fields is
\eql{sphi}{S_{\phi}=\int d^4x\sqrt{|g|}\left[-g^{\mu\nu}D_\mu\phi D_\nu\bar{\phi}-iV^\mu\left(\bar{\phi}D_\mu\phi-\phi D_\mu\bar{\phi}\right)\right],}
where 
\eq{D_\mu\phi=\left(\dm-iqA_\mu\right)\phi,\quad D_\mu\bar{\phi}=\left(\dm+iqA_\mu\right)\bar{\phi}.}
We now expand $\phi$ in Kaluza-Klein (KK) modes around the $S^1$ by defining
\eq{\phi=\sum_k\phi_ke^{ik\te},\quad \phi_k=\int \frac{d\te}{2\pi}\phi e^{-ik\te}.}
In terms of these modes, the action \eqref{sphi} on $AdS_3$ takes the form
\begin{equation}\begin{gathered}S_{\phi}=-2\pi R\sum_k\int d^3x_{AdS}\sqrt{|g_{AdS}|}\left[g^{ij}\partial_i\phi_k\partial_j\bar{\phi}_k+m_k^2\phi_k\bar{\phi}_k\right],\\ m_k^2=\frac{k^2}{R^2}+\frac{q(q-2)}{L^2}+\frac{2k(q-1)}{RL}.
\end{gathered}\label{effmass}
\end{equation}
The asymptotic solution near the boundary is given by the standard formula
\eq{\phi_k=\phi_{k_+}(x,t)r^{\Delta_{k_+}}+\phi_{k_-}(x,t)r^{\Delta_{k_-}},}
with \eql{Sdimension}{\Delta_{k_\pm}=1\pm\sqrt{1+L^2m_k^2}=1\pm\left(\frac{kL}{R}+q-1\right).}
The physical modes must have $\Delta>0$. We will fix the other modes on the boundary and find the correct boundary action to yield a well-defined variational principle. 

The variation of the bulk action is
\eq{\delta S_{\phi}=&2\pi R\sum_k\int d^3x_{AdS}\delta\bar{\phi}_k\left[\partial_i\sqrt{|g|}g^{ij}\partial_j-\sqrt{|g|}m_k^2 \right]\phi_k\\&+2\pi R\sum_k\int d^3x_{AdS}\delta\phi_k\left[\partial_i\sqrt{|g|}g^{ij}\partial_j-\sqrt{|g|}m_k^2 \right]\bar{\phi}_k\\
&-2\pi R\sum_k\int d^2x \sqrt{|h|}n_i g^{ij}\left[\partial_j\bar{\phi}_k\delta\phi_k+\partial_j\phi_k\delta\bar{\phi}_k\right],}
where $h$ is the induced metric on the boudnary of $AdS$ space (say with some cutoff on the radial direction), $d^2x$ denotes an integration over the variables $x$ and $t$ which label this boundary, and $n$ is a vector transverse to the boundary.
The bulk terms vanish on shell. In order for the boundary term to vanish, we add a boundary action $S_{bndy,\phi}$ to $S_\phi$ such that the total variation vanishes when the physical boundary conditions are held. 
If we choose 
\eq{S_{bndy,\phi+}&=\int d^2xd\te\sqrt{|h|}\left[n_\mu \partial^\mu(\phi\bar{\phi})-\frac{i}{2R}\phi D_\te \bar{\phi}+\frac{i}{2R}\bar{\phi}D_\te\phi\right],}
we get the variation
\eq{\delta S_{\phi}+\delta& S_{bndy,\phi+}=\\&
\int d^2xd\te\sqrt{|h|}\left[\phi\left(n_\mu g^{\mu\nu}\dn-\frac{i}{R}D_\te\right)\delta\bar{\phi}+\bar{\phi}\left(n_\mu g^{\mu\nu}\dn+\frac{i}{R}D_\te\right)\delta\phi\right].}
The operators $\left(n_\mu g^{\mu\nu}\dn\pm\frac{i}{R}D_\te\right)$ annihilate $\phi_{k_+},\ \bar{\phi}_{k_+}$ for every $k$. Therefore, by fixing the $\phi_{k-}$ modes, $\delta\phi_{k_-}=\delta\bar{\phi}_{k_-}=0$, the variation of the action vanishes.
If, on the other hand, we take the boundary action to be
\eq{S_{bndy,\phi-}&=\int d^2xd\te\sqrt{|h|}\left[n_\mu \partial^\mu(\phi\bar{\phi})+\frac{i}{2R}\phi D_\te \bar{\phi}-\frac{i}{2R}\bar{\phi}D_\te\phi-\frac{2}{L}\phi\bar{\phi}\right],}
the variation becomes
\eq{\delta S_{\phi}+&\delta S_{bndy,\phi-}=\\
&\int d^2xd\te\sqrt{|h|}\left[\phi\left(n_\mu g^{\mu\nu}\dn-\frac{2}{L}+\frac{i}{R}D_\te\right)\delta\bar{\phi}+\bar{\phi}\left(n_\mu g^{\mu\nu}\dn-\frac{2}{L}-\frac{i}{R}D_\te\right)\delta\phi \right].}
In this case, the variation vanishes when fixing $\delta\phi_{k_+}=\delta\bar{\phi}_{k_+}=0$. 

In order to satisfy $\Delta>0$ for all the fluctuating modes, we need to have a mixed $k$-dependent boundary action. For the KK modes with $0 < \Delta < 2$ we have two possibilities for the boundary action, while otherwise we have just one choice.
The full boundary action should then be
\eq{S_{bndy,\phi}=\sum_{k=k^*}^{\infty}S_{bndy,\phi,k+}+\sum_{k=-\infty}^{k^*-1}S_{bndy,\phi,k-},} 
with $k^*$ such that for $k<k^*$, $\Delta_{k_-}>0$ and for $k\geq k^*$, $\Delta_{k_+}>0$ (in some cases there may be more than
one possible choice for this $k^*$, giving different theories on $AdS_3\times S^1$).

\subsubsection{Fermion}

The bulk action for a fermion in a chiral multiplet is a kinetic term and a term that couples the fermions to the background field,
\eq{&S_{\psi}=\int d^4x\sqrt{|g|}\left[-i\bar{\psi}\bar{\sigma}^\mu \left(D_\mu-\frac{i}{2}V_\mu\right)\psi\right],\\ &D_\mu\psi=\left(\nabla_\mu-i(q-1)A_\mu\right)\psi.}
Here we took the fermion to have R-charge $(q-1)$, consistent with putting it in the same multiplet as the scalar with R-charge $q$.

As in the case of the scalar, we expand in KK modes 
\eq{\psi=\sum_k\psi_ke^{ik\te},\quad \psi_k=\int \frac{d\te}{2\pi}\psi e^{-ik\te},}
and solve the equations of motion near the boundary. We take the fermion to be periodic around the circle, anticipating that this will be required for preserving supersymmetry. 

The asymptotic solution is given by
\eql{Fdimension}{&\psi_k=\psi_+\mat{1\\1}r^{\Delta_{f,k_+}}+\psi_-\mat{1\\-1}r^{\Delta_{f,k_-}},\\ &\Delta_{f,k_\pm}=1\pm\left(\frac{kL}{R}+q-\onov{2}\right).}
As explained in section \ref{algebra}, the $2d$ conformal dimensions of $\psi_\pm$ satisfy $h_L-h_R=\pm\onov{2}$.  The spectrum can be written as
\eq{\left(h_L,h_R\right)_{k,+}=\left(\onov{2}+\frac{kL}{2R}+\frac{q}{2},\frac{kL}{2R}+\frac{q}{2}\right)\ ,\quad \left(h_L,h_R\right)_{k,-}=\left(\onov{2}-\frac{kL}{2R}-\frac{q}{2},1-\frac{kL}{2R}-\frac{q}{2}\right)\ .}

As before, in order to have a well defined variational principle, we need to split our boundary action. If we take the boundary action to be
\eq{S_{bndy,\psi}=\sum_{k=k^*}^{\infty}S_{bndy,\psi,k+}+\sum_{k=-\infty}^{k^*-1}S_{bndy,\psi,k-}\quad,} 
with 
\eq{S_{bndy,\psi\pm}=\frac{1}{2}\int d^2xd\te\sqrt{|h|}\bar{\psi}\left[in_\mu\bar{\sigma}^\mu\pm R\bar{\sigma}^\te\right]\psi\quad,}
then $\delta S_\psi+\delta S_{bndy,\psi\pm}=0$, when we keep the modes $\psi_{k,\pm}$.

Note that because $\Delta_{f,k_\pm}=\Delta_{s,k_\pm}\pm\onov{2}$, the constraints on $k^*$ for the fermion are the same as we found for the scalar. This is of course important for SUSY, as will be shown in the next section.

\subsubsection{Gauge field}
\label{Vec}

The action for a free $U(1)$ gauge field $v_{\mu}$ is the Maxwell term 
\eql{maxwell}{S_{YM}=-\onov{4}\int d^4x\sqrt{|g|}F_{\mu\nu}F^{\mu\nu}.}
When expanded in KK modes, 
\eq{v_{\mu}=\sum_k v_{\mu}^{(k)} e^{ik\te},\quad v_{\mu}^{(k)}=\int \frac{d\te}{2\pi} v_{\mu} e^{-ik\te},}
it takes the following form
 \eq{S_{YM}=-\pi R\sum_k\int d^3x_{AdS}\sqrt{|g_{AdS}|}\left[\onov{2}F_{ij}^{(k)}F^{ij(-k)}+\frac{k^2}{R^2}v_i^{(k)}v^{i(-k)}-\frac{2ik}{R}v^{i(k)}\partial_iv^{\te(-k)}+\partial_iv_\te^{(k)}\partial^iv^{\te(-k)} \right],}
where $i,j$ go over the $AdS_3$ coordinates.

 We can choose a gauge where $v_\te^{(k\neq0)}=0,\ v_r^{(k=0)}=0$, for which the action simplifies to
 \eq{S_{YM}=-\pi R\int d^3x_{AdS}\sqrt{|g_{AdS}|}\left[\sum_k \left( \onov{2} F_{ij}^{(k)}F^{ij(-k)}+\frac{k^2}{R^2}v_i^{(k)}v^{i(-k)}\right) +\partial_iv_\te^{(0)}\partial^iv^{\te(0)} \right].}
 For the $k=0$ KK mode, the matter content consists of a massless scalar $v_\te^{(0)}$ and a $3d$ massless gauge field $v_i^{(0)}$. The normalizable modes are a scalar whose dimension in the $2d$ conformal algebra is $\Delta=2$, and a $U(1)$ gauge field on the boundary, while the non-normalizable modes that couple to them are a scalar of dimension $\Delta=0$ and a conserved current. Note that from the point of view of the $2d$ superconformal algebra this means that (for the action \eqref{maxwell}) we do not get a conserved current representation, but rather a representation that contains the $2d$ field strength arising from the value of $v_{\mu}$ on the boundary. For the $k\neq0$ modes, we have a complex massive vector field $v_i^{(k)}$.

The asymptotic solution to the equations of motion gives the following dimensions for the $k$'th KK mode of $v_i$,
\eq{\Delta_k =\pm\frac{kL}{R}.}
Similar to the previous cases, we demand that the variation of the total action should vanish when the physical boundary conditions are satisfied. The boundary action that we need to add is
\eq{S_{bndy,v}=\sum_{k=-\infty}^{-1}S^-_{-k}+S^0+\sum_{k=1}^{\infty}S^+_{k},}
 where
 \eq{S_k^{\pm}=\onov{2}\int d^2xd\te\sqrt{|h|}\left[n_\mu v_a^{(k)} F^{(-k) \mu a}\pm iR v_a^{(k)} F^{(-k) \te a}\right],\quad S^0=\int d^2xd\te\sqrt{|h|}n_\mu F^{(0)\mu a} v_a^{(0)}.}
The generalization to non-Abelian gauge fields is straightforward.

\subsection{A free $\mathcal{N}=1$ chiral multiplet}

The on-shell supersymmetric chiral multiplet consists a complex scalar $\phi$ and a Weyl fermion $\psi$. Following the previous subsections, 
the action for the free massless chiral multiplet with R-charge $q$ on $AdS_3\times S^1$ is given by
\eq{S_{chiral}=S_{chiral,bulk}+S_{bndy,\phi}+S_{bndy,\psi},}
where
\eq{S_{chiral,bulk}=\int d^4x\sqrt{|g|}\left[-i\bar{\psi}\bar{\sigma}^\mu D_\mu\psi-g^{\mu\nu}D_\mu\phi D_\nu\bar{\phi}-iV^\mu\left(\bar{\phi}D_\mu\phi-\phi D_\mu\bar{\phi}\right)\right],}
and the covariant derivatives are 
\eq{D_\mu\phi=(\dm-iqA_\mu)\phi,\quad D_\mu\psi=\left(\nabla_\mu-i(q-1)A_\mu-\frac{i}{2}V_\mu\right)\psi.}
The SUSY variations of the fields are
\eq{\delta\phi=-\sqrt{2}\zeta\psi,\quad \delta\bar{\phi}=-\sqrt{2}\zeb\bar{\psi},\quad \delta\psi=-\sqrt{2}i\sigma^\nu\zeb D_\nu\phi,\quad \delta\bar{\psi}=\sqrt{2}i\zeta\sigma^\nu D_\nu\bar{\phi}.}

The full action $S_{chiral}$ accompanied with the boundary conditions specified in section \ref{N0} is invariant under all four supercharges.
Each scalar $k_\pm$ mode with dimension \eqref{Sdimension} has a superpartner fermion $k_\pm$ mode with dimension \eqref{Fdimension}, such that 
\eq{\Delta_{f,k_\pm}=\Delta_{s,k_\pm}\pm\onov{2},\qquad s_{f,k\pm}=\pm\onov{2}.}
By comparing to the known $\mathcal{N}=(0,2)$ multiplets, we see that the $(\phi_{k_+},\psi_{k_+})$ form chiral multiplets, and $(\phi_{k_-},\psi_{k_-})$ form Fermi multiplets.

\subsubsection{BF bound saturation}

According to Breitenlohner-Freedman  \citep{breitenlohner1982stability,breitenlohner1982positive},
the minimal mass of a scalar on $AdS_{3}$ can be
\eq{m^2_{BF}=-\onov{L^2},}
otherwise we get complex dimensions from the point of view of the $2d$ conformal algebra. Supersymmetry guarantees
that $m^2 \geq m_{BF}^2$ (see \eqref{effmass}), but we should discuss the special case where this bound is saturated.
This happens if there exists an integer $k$ such that $\frac{kL}{R}+q=1$, and then for this $k$ $\Delta_{k\pm}$ coincide. In this case, the asymptotic solution to the Klein-Gordon equation is
\eq{\phi=\phi_+r+\phi_-r\log(r).}
If we fix the non-normalizable mode $\phi_-$ on the boundary, the analysis is similar to the one done in the previous sections and all supercharges are preserved. The other boundary condition breaks the conformal symmetry.

\subsubsection{A free massive chiral multiplet}

We can add a mass in a supersymmetric way by adding a superpotential $W=\onov{2} m\phi^2$ and taking the $R$-charge of the scalar to be $q=1$. After integrating out the auxilliary field $F$, we get the bulk Lagrangian
\eq{\lag=-i\bar{\psi}\bar{\sigma}^\mu \left(\nabla_\mu-\frac{i}{2}V_\mu\right)\psi-g^{\mu\nu}D_\mu\phi D_\nu\bar{\phi}-iV^\mu\left(\bar{\phi}D_\mu\phi-\phi D_\mu\bar{\phi}\right)-m^2\phi\bar{\phi}-\onov{2}m\psi\psi-\onov{2}m\bar{\psi}\bar{\psi}.}
The spectrum is modified due to the mass. The equation of motion of the scalar is
\eq{D_\mu\sqrt{|g|}g^{\mu\nu}(D_\nu+2iV_\nu)\phi-\sqrt{|g|}m^2\phi=0.}
The dimension of the $k$'th KK mode is given by
\eq{\Delta_{k\pm}=1\pm\sqrt{m^2L^2+\frac{k^2L^2}{R^2}}.}

For the Fermion, we have the coupled equations
\eq{\bar{\sigma}^\mu\left(\nabla_\mu-\frac{i}{2}V_\mu\right)\psi_k=im\bar{\psi}_{-k},\quad \sigma^\mu\left(\nabla_\mu+\frac{i}{2}V_\mu\right)\bar{\psi}_{-k}=im\psi_k.}
Asymptotically, we get
\eq{\mat{\frac{k}{R}+\onov{2L}&\frac{1-\Delta}{L}\\\frac{\Delta-1}{L}&-\frac{k}{R}-\onov{2L}}\psi_k=-m\bar{\psi}_{-k},\quad \mat{-\frac{k}{R}+\onov{2L}&\frac{\Delta-1}{L}\\\frac{1-\Delta}{L}&\frac{k}{R}-\onov{2L}}\bar{\psi}_{-k}=-m\psi_{k}.}
Plugging one into the other, we get
\eq{\mat{\onov{4L^2}-\frac{k^2}{R^2}+\frac{(\Delta-1)^2}{L^2}-m^2&\frac{1-\Delta}{L^2}\\\frac{1-\Delta}{L^2}&\onov{4L^2}-\frac{k^2}{R^2}+\frac{(\Delta-1)^2}{L^2}-m^2}\psi_k=0,}
and the dimensions
\eq{\Delta_{1\pm}=\frac{1}{2}\pm\sqrt{m^2L^2+\frac{k^2L^2}{R^2}}\ ,\  \Delta_{2\pm}=\frac{3}{2}\pm\sqrt{m^2L^2+\frac{k^2L^2}{R^2}},}
with the asymptotic expansions
\eq{&\psi_k=\psi_{1+}\mat{1\\-1}r^{\Delta_{1+}}+\psi_{2+}\mat{1\\1}r^{\Delta_{2+}}+\psi_{1-}\mat{1\\-1}r^{\Delta_{1-}}+\psi_{2-}\mat{1\\1}r^{\Delta_{2-}},\\
&\bar{\psi}_{-k}=\bar{\psi}_{1+}\mat{1\\1}r^{\Delta_{1+}}+\bar{\psi}_{2+}\mat{1\\-1}r^{\Delta_{2+}}+\bar{\psi}_{1-}\mat{1\\1}r^{\Delta_{1-}}+\bar{\psi}_{2-}\mat{1\\-1}r^{\Delta_{2-}},}
where
\eq{\bar{\psi}_{1,2\pm}=-\left(\frac{k}{Rm}\pm\sqrt{1+\frac{k^2}{R^2m^2}}\right)\psi_{1,2\pm}.}
Now the SUSY transformations mix the fields $\phi_k,\ \bar{\phi}_{-k},\ \psi_k,\ \bar{\psi}_{-k}$ but they can be diagonalized such that they split into four multiplets with dimensions
\eq{&\Delta_s=1+\sqrt{m^2L^2+\frac{k^2L^2}{R^2}}\ , \ \Delta_f=\frac{3}{2}+\sqrt{m^2L^2+\frac{k^2L^2}{R^2}}\ ,\ s_f=\onov{2}\ ,\ Chiral, \\
&\Delta_s=1+\sqrt{m^2L^2+\frac{k^2L^2}{R^2}}\ , \ \Delta_f=\frac{1}{2}+\sqrt{m^2L^2+\frac{k^2L^2}{R^2}}\ ,\ s_f=-\onov{2}\ ,\ Fermi, \\
&\Delta_s=1-\sqrt{m^2L^2+\frac{k^2L^2}{R^2}}\ , \ \Delta_f=\frac{3}{2}-\sqrt{m^2L^2+\frac{k^2L^2}{R^2}}\ ,\ s_f=\onov{2}\ ,\ Chiral, \\
&\Delta_s=1-\sqrt{m^2L^2+\frac{k^2L^2}{R^2}}\ , \ \Delta_f=\frac{1}{2}-\sqrt{m^2L^2+\frac{k^2L^2}{R^2}}\ ,\ s_f=-\onov{2}\ ,\ Fermi .} 
The boundary conditions fix two of them -- either the first and fourth, or the second and third -- such that only two give operators in the $2d$ superconformal algebra, one chiral and one Fermi multiplet for every $k$.

\subsubsection{Breaking the $S^1$ isometry}

As was discussed in section \ref{SCalgebra}, in order to have supersymmetry on $AdS_p\times S^q$, some mixture of the $R$-symmetry and the $S^q$ isometries that appears on the right-hand side of the supercharges anti-commutator must be preserved. Specifically, for the case of $\mathcal{N}=1$ on $AdS_3\times S^1$, from the algebraic analysis we know that we must preserve a specific combination of the $S^1$ isometry and the $U(1)_R$ generator, but can break each one of them separately. 
A simple field theory realization for this is to add a $\te$-dependent mass term to the free chiral multiplet. This is done by adding the superpotential
\eq{W=me^{2i(q-1)A_\mu x^\mu}\phi^2\ ,\ \bar{W}=me^{-2i(q-1)A_\mu x^\mu}\bar{\phi}^2.}
By taking $q=1$ we get the regular massive chiral multiplet discussed in the previous section, but we will keep $q$ arbitrary such that the Lagrangian
\eq{\lag=-i\bar{\psi}\bar{\sigma}^\mu D_\mu\psi-g^{\mu\nu}D_\mu\phi D_\nu\bar{\phi}-2iV^\mu\bar{\phi}D_\mu\phi-m^2|\phi|^2-\frac{m}{2}e^{2i(q-1)A_\mu x^\mu}\psi\psi-\frac{m}{2}e^{-2i(q-1)A_\mu x^\mu}\bar{\psi}\bar{\psi}}
breaks the $S^1$ isometry and the $U(1)_R$ symmetry.
 The theory is still supersymmetric as before, with the dimensions modified to
 \eq{&\Delta_{\phi}=1\pm\sqrt{m^2L^2+\left(\frac{kL}{R}+q-1\right)^2},\\
 &\Delta_{\psi}=\onov{2}\pm\sqrt{m^2L^2+\left(\frac{kL}{R}+q-1\right)^2}\ ,\ \frac{3}{2}\pm\sqrt{m^2L^2+\left(\frac{kL}{R}+q-1\right)^2},}
but it no longer has an extra $U(1)$ global symmetry.

\subsection{A free $\mathcal{N}=1$ vector multiplet}
\label{VecMultiplet}

The bulk action for a $U(1)$ gauge multiplet is
\eq{S_{vector}=\int d^4x\sqrt{|g|}\left[-\onov{4}F_{\mu\nu}F^{\mu\nu}-i\bar{\la}\bar{\sigma}^\mu\left(\nabla_\mu+\frac{i}{2}A_\mu\right)\la \right],}
and the transformation rules are
\eq{\delta v_\mu=i\zeta\sigma_\mu\bar{\la}-i\la\sigma^\mu\zeb,\quad \delta\la=F_{\mu\nu}\sigma^{\mu\nu}\zeta.}

   The KK modes are defined as above, and we can compute their variations. We will do it explicitly for $\delta_{b}, \delta_{\bar{b}}$, which is enough to understand to structure of the multiplet.
   Some of the transformations are:
   \begin{center}

\begin{tabular}{cccc}
 \\[0.1pt]
\multirow{3}{*}{$k=0,+$}&$\delta_b\la=\onov{LR}(\dt-\dx)v_{\te}+\frac{i}{L^2}(v_{x}-v_{t})$\\
&$\delta_{\bar{b}}v_{\te}=iR\la$\\
&$\delta_{\bar{b}}(v_{x}-v_{t})=L(\dt-\dx)\la$
 \\[0.1pt]
\hline
 \\[0.1pt]
\multirow{3}{*}{$k=0,-$}&$\delta_b\la=\onov{L^2}F_{tx}-\frac{2i}{LR}v_{\te}$\\
&$\delta_{\bar{b}}v_{\te}=\frac{R}{2}(\dx-\dt)\la$\\
&$\delta_{\bar{b}}F_{tx}=iL(\dt-\dx)\la$
 \\[0.1pt]
 \hline
 \\[0.1pt]
 \multirow{2}{*}{$k\neq 0,+$}&$\delta_b\la=\frac{2ik}{LR}(v_{x}-v_{t})$\\
&$\delta_{\bar{b}}(v_{x}-v_{t})=\frac{R}{k}(\dt-\dx)\la$
 \\[0.1pt]
\hline
 \\[0.1pt]
\multirow{2}{*}{$k\neq0,-$}&$\delta_b\la=\onov{L^2}(\dt-\dx)(v_{x}+v_{t})$\\
&$\delta_{\bar{b}}(v_{x}+v_{t})=-2iL\la$.
 \\[0.1pt]
\end{tabular}   
   
   \end{center}   
   
For $k=0$, the $-$ modes are the physical ones. They contain a $2d$ gauge field $v_{i}$, a dimension $\left(\onov{2},1\right)$ fermion $\la_-$ and a dimension $\left(1,1\right)$ scalar $v_{\te}$.
Its dual multiplet (containing the couplings to these operators) contains a conserved current, a dimension $\left(\onov{2},0\right)$ fermion and a dimension $0$ scalar.

 For $k>0$, the $+$ modes are physical. They contain a vector of dimension $\left(1+\frac{kL}{2R},\frac{kL}{2R}\right)$, and a fermion of dimension $\left(\onov{2}+\frac{kL}{2R},\frac{kL}{2R}\right)$.

 For $k<0$, the $-$ modes are physical. They contain a vector of dimension $\left(-\frac{kL}{2R},1-\frac{kL}{2R}\right)$ and a fermion of dimension $\left(\frac{1}{2}-\frac{kL}{2R},1-\frac{kL}{2R}\right)$.

\subsection{A free $\mathcal{N}=2$ hypermultiplet in the $\mathcal{N}=(0,4)$ case}

For $\mathcal{N}=2$ theories we can either just
guess the form of the Killing spinors and supersymmetric actions, based on the results for the $\mathcal{N}=1$ case and on
requiring their consistency with the full $\mathcal{N}=2$ supersymmetry, or derive them
by coupling to a background $\mathcal{N}=2$ superconformal gravity, as described in appendix \ref{KSE}.

We begin in this subsection by studying the $\mathcal{N}=2$ hypermultiplet using the supersymmetry that gives a $\mathcal{N}=(0,4)$ superconformal algebra. Denoting the scalar fields by $\phi_{1,2}$ and the fermions by $\psi_{1,2}$, the supersymmetry transformations of the different fields are
\eq{&\delta\mat{\phi_1\\\bar{\phi}_2}=-\sqrt{2}\psi_1\mat{\zeta_1\\\zeta_2}-\sqrt{2}\bar{\psi}_2\mat{\zeb_2\\-\zeb_1},\\
&\delta\psi_1=-\sqrt{2}i\sigma^\nu\mat{\zeb_1&\zeb_2}(D_\nu-iV_\nu)\mat{\phi_1\\\bar{\phi}_2},\\
&\delta\bar{\psi}_2=\sqrt{2}i\bar{\sigma}^\nu\mat{-\zeta_2&\zeta_1}(D_\nu+iV_\nu)\mat{\phi_1\\\bar{\phi}_2}.}
Here $V_{\mu}$ is the same as in the $\mathcal{N}=1$ case above.

The bulk action, equations of motion and Killing spinor equations are
\eq{\lag=-g^{\mu\nu}D_\mu\phi_i D_\nu\bar{\phi}_i+V^\mu V_\mu \phi_i\bar{\phi}_i-i\bar{\psi}_i\bar{\sigma}^\mu\left(D_\mu-\frac{i}{2}V_\mu\right)\psi_i,}
\eq{&(D^2+V^2)\phi_i=(D^2+V^2)\bar{\phi}_i=\bar{\sigma}^\mu\left(D_\mu-\frac{i}{2}V_\mu\right)\psi_i=\sigma^\mu\left(D_\mu+\frac{i}{2}V_\mu\right)\bar{\psi}_i=0,\\
&\qquad\qquad\qquad\qquad\qquad  D_\mu\zeta_i=-iV^\rho\sigma_{\mu\rho}\zeta_i,\ D_\mu\zeb_i=iV^\rho\bar{\sigma}_{\mu\rho}\zeb_i,}
with the covariant derivatives defined as
\eq{&D_\mu\phi_1=(\dm-iqA_\mu)\phi_1\ ,\ D_\mu\bar{\phi}_2=(\dm-iqA_\mu)\bar{\phi}_2\ ,\ D_\mu\psi_1=(\nabla_\mu-i(q-1)A_\mu)\psi_1,\\
& D_\mu\bar{\psi}_2=(\nabla_\mu+i(q+1)A_\mu)\bar{\psi}_2\ ,\ D_\mu\zeta_i=(\nabla_\mu-iA_\mu)\zeta_i\ ,\ D_\mu\zeb_i=(\nabla_\mu+iA_\mu)\zeb_i,}
and their complex conjugates.
$A_{\mu}$ here is a background field for the $U(1)$ $R$-symmetry of the $\mathcal{N}=2$ theory; this symmetry must exist in order to use this construction, based on our algebraic discussion. We take the $R$-charge of the scalars in the hypermultiplet to be $q$. As in our discussion of the $\mathcal{N}=1$ case in section \ref{algebra}, we can perform shifts in the background value of $A_{\mu}$ corresponding to large gauge transformations. In our conventions of this section, in order for the Killing spinors to be periodic, we need to perform a specific shift which amounts to setting $A_{\mu}=0$, and we will use this value below. The $\mu=\te$ part of the Killing spinor equations here seems different than the one we used for the $\mathcal{N}=1$ case. This happened because here we used the conformal supergravity approach, rather than coupling to regular supergravity. As discussed in the introduction, the two approaches must give the same answer. This is consistent because the $\mathcal{N}=1$ Killing spinor equations can be brought to the form we found here by redefinitions of the background fields that have no effect on the physics (they still describe $AdS_3\times S^1$ with the same parameters). 

The solution to the Killing spinor equation is
\eq{\zeta_i=a_i r^{\onov{2}}\mat{1\\1}+(b_i+ia_iz) r^{-\onov{2}}\mat{1\\-1}.}

The asymptotic dimensions of the fields are
\eq{\Delta_{\phi,\pm}=1\pm \frac{kL}{R},\ \Delta_{\psi,\pm}=1\pm \left(\frac{kL}{R}+\onov{2} \right).}

Each representation is characterized by the central charge $-iD_\te=k$. At first sight this is confusing because in our discussion of section \ref{d4N2} we found that the central charge was a combination of the KK momentum and the $U(1)_R$ charge -- this is related to the conventions we used above for defining $A_{\mu}$, which are equivalent to shifting the KK momentum by the R-charge. The supersymmetry algebra that we find from the transformations above is precisely the $\mathcal{N}=(0,4)$ supersymmetry algebra with this central charge, as in section \ref{d4N2}. We have two kinds of representations, the $+$ and the $-$. The computations are the same as in the $\mathcal{N}=(0,2)$ case. The $+$ representations form  a $\mathcal{N}=(0,4)$ hypermultiplet, and the $-$ representations form a $\mathcal{N}=(0,4)$ Fermi multiplet.
In particular,
the scalars are charged under the $SU(2)_R$ symmetry in the $\mathcal{N}=(0,4)$ superconformal algebra, as can be shown from\footnote{From the Killing spinors, we can construct $V_\mu\zeta_i\sigma^\mu\zeb_j=\frac{2}{L}\left(a_i\bar{b}_j-b_i\bar{a}_j\right)$ which is coordinate independent, and is therefore a good $R$-symmetry generator. }
\eq{\delta^2\phi_1&=2i\zeta_i\sigma^\nu\zeb_i D_\nu\phi_1+2V_\nu\zeta_1\sigma^\nu\zeb_1\phi_1-2V_\nu\zeta_2\sigma^\nu\zeb_2\phi_1+4V_\nu\zeta_1\sigma^\nu\zeb_2\bar{\phi}_2,\\
\delta^2\bar{\phi}_2&=2i\zeta_i\sigma^\nu\zeb_i D_\nu\bar{\phi}_2-2V_\nu\zeta_1\sigma^\nu\zeb_1\bar{\phi}_2+2V_\nu\zeta_2\sigma^\nu\zeb_2\bar{\phi}_2+4V_\nu\zeta_2\sigma^\nu\zeb_1\phi_1.}

\subsection{A free $\mathcal{N}=2$ hypermultiplet in the $\mathcal{N}=(2,2)$ case}

In this section we discuss the same theory when we put it on $AdS_3\times S^1$ with the $\mathcal{N}=(2,2)$ superconformal algebra. As opposed to the $\mathcal{N}=(0,4)$ case, here the Killing spinors can have a relative phase (i.e. different $\te$-dependence). As discussed in appendix \ref{KSE}, we will work with $\te$-independent Killing spinors, such that the SUSY transformations don't mix different KK modes and we can have a supersymmetric theory for every ratio $\frac{R}{L}$. For that, we choose $A_\mu=0$ and the Killing spinor equations for two spinors of opposite chirality are
\eq{\partial_\mu\zeta_1=-iV^\rho\sigma_{\mu\rho}\zeta_1,\ \partial_\mu\zeta_2=+iV^\rho\sigma_{\mu\rho}\zeta_2.}
The Lagarangian and the equations of motion are
\eq{&\lag=-g^{\mu\nu}\dm\phi_i\dn\bar{\phi}_i+V^2|\phi_1|^2-2iV^\mu\phi_2\dm\bar{\phi}_2-i\bar{\psi}_1\bar{\sigma}^\mu\left(\nabla_\mu-\frac{i}{2}V_\mu\right)\psi_1-i\bar{\psi}_2\bar{\sigma}^\mu\left(\nabla_\mu+\frac{i}{2}V_\mu\right)\psi_2\ ,\\&\bar{\sigma}^\mu(\nabla_\mu-\frac{i}{2}V_\mu)\psi_1=\bar{\sigma}^\mu(\nabla_\mu+\frac{i}{2}V_\mu)\psi_2=0\ ,\ \left[\Box+V^2\right]\bar{\phi}_1=\left[\Box-2iV^\mu\dm\right]\bar{\phi}_2=0,}
with the transformation rules
\eq{&\delta\phi_1=-\sqrt{2}\zeta_1\psi_1-\sqrt{2}\zeb_2\bar{\psi}_2,\\
&\delta\bar{\phi}_2=-\sqrt{2}\psi_1\zeta_2+\sqrt{2}\zeb_1\bar{\psi}_2,\\
&\delta\psi_1=-\sqrt{2}i\sigma^\nu\zeb_1(\partial_\nu-iV_\nu)\phi_1-\sqrt{2}i\sigma^\nu\zeb_2 \partial_\nu\bar{\phi}_2,\\
&\delta\bar{\psi}_2=-\sqrt{2}i\bar{\sigma}^\nu\zeta_2(\partial_\nu-iV_\nu)\phi_1+\sqrt{2}i\bar{\sigma}^\nu\zeta_1\partial_\nu\bar{\phi}_2.}
 
The asymptotic dimensions of the fields are
\eql{dimphipsi}{&\Delta_{\phi_1}=1\pm\frac{kL}{R},\ \Delta_{\phi_2}=1\pm\left(\frac{kL}{R}-1\right),\\& \Delta_{\psi_1}=1\pm\left(\frac{kL}{R}+\onov{2} \right),\ \Delta_{\psi_2}=1\pm\left(\frac{kL}{R}-\onov{2}\right).}
The modes that sit in the same multiplets are
\eq{k_{\phi_1}=k_{\psi_1}=k,\ k_{\phi_2}=k_{\psi_2}=-k.}
Using \eqref{dimphipsi}, we find that in the same multiplet we have
\eq{&\Delta_{\phi_1,+}=\Delta_{\psi_1,+}-\onov{2}=\Delta_{\psi_2,-}-\onov{2}=\Delta_{\phi_2,-}-1,\\
& \Delta_{\phi_1,-}=\Delta_{\psi_1,-}+\onov{2}=\Delta_{\psi_2,+}+\onov{2}=\Delta_{\phi_2,+}+1.}
 We see that we can have a consistent supersymmetric theory for any value of $\frac{R}{L}$.
 For the two signs we find either an $\mathcal{N}=(2,2)$ multiplet that is made out of a $\mathcal{N}=(0,2)$ chiral multiplet and a $\mathcal{N}=(2,0)$ Fermi multiplet, or vice versa.

\subsection{A free $\mathcal{N}=2$ vector multiplet in the $\mathcal{N}=(0,4)$ case}

The Killing spinor equation and the equation of motion for the two fermions and the scalar are 
\eq{D_\mu\zeta_i+iV^\rho\sigma_{\mu\rho}\zeta_i&=0,\\
\sigma^\mu\left(D_\mu-\frac{i}{2}V_\mu\right)\bar{\psi}&=0,\\\sigma^\mu\left(D_\mu-\frac{i}{2}V_\mu\right)\bar{\la}&=0,\\
\left[D^2+2iV^{\mu} D_{\mu}\right]\phi&=0.}
The bulk action is
\eq{\lag=-\onov{4}F_{\mu\nu}F^{\mu\nu}-i\bar{\la}\bar{\sigma}^\mu\left(\nabla_\mu-\frac{i}{2}V_\mu\right)\la-i\bar{\psi}\bar{\sigma}^\mu\left(\nabla_\mu-\frac{i}{2}V_\mu\right)\psi-g^{\mu\nu}\dm\phi\dn\bar{\phi}-2iV^\mu\phi\dm\bar{\phi}.}
The SUSY variations are
\eq{\delta v_\mu=& -i\zeta_1\sigma_\mu\bar{\la}+i\la\sigma_\mu\zeb_1-i\zeta_2\sigma_\mu\bar{\psi}+i\psi\sigma_\mu\zeb_2, \\
\delta \la=&F_{\mu\nu}\sigma^{\mu\nu}\zeta_1+\sqrt{2}i\sigma^\nu\zeb_2D_\nu\phi,\\
\delta \psi=&F_{\mu\nu}\sigma^{\mu\nu}\zeta_2-\sqrt{2}i\sigma^\nu\zeb_1D_\nu\phi,\\
\delta\bar{\la}=&-F_{\mu\nu}\zeb_1\bar{\sigma}^{\mu\nu}+\sqrt{2}i\zeta_2\sigma^\nu D_\nu\bar{\phi},\\
\delta\bar{\psi}=&F_{\mu\nu}\zeb_2\bar{\sigma}^{\mu\nu}-\sqrt{2}i\zeta_1\sigma^\nu D_\nu\bar{\phi},\\
\delta\phi=&-\sqrt{2}\zeta_1\psi+\sqrt{2}\zeta_2\la,\\
\delta\bar{\phi}=&-\sqrt{2}\bar{\psi}\zeb_1+\sqrt{2}\bar{\la}\zeb_2.}
In this case the fermions are charged under $SU(2)_R$, as can be seen for example from
\eq{\delta_{1\bar{2}}^2\la=4V_\mu\zeta_1\sigma^\mu\zeb_2 \psi.}

The dimensions of the scalar and fermions are
\eq{\Delta_{\phi}=1\pm\left(\frac{kL}{R}-1\right),\Delta_{\psi}=\Delta_{\la}=1\pm\left(\frac{kL}{R}-\onov{2}\right).}
This $\mathcal{N}=(0,4)$ multiplet is the combination of $\mathcal{N}=(0,2)$ chiral and vector multiplets.

\subsection{A free $\mathcal{N}=2$ vector multiplet in the $\mathcal{N}=(2,2)$ case}

 The transformations in this case are
\eq{\delta v_\mu=& -i\zeta_1\sigma_\mu\bar{\la}+i\la\sigma_\mu\zeb_1-i\zeta_2\sigma_\mu\bar{\psi}+i\psi\sigma_\mu\zeb_2, \\
\delta \la=&F_{\mu\nu}\sigma^{\mu\nu}\zeta_1+\sqrt{2}i\sigma^\nu\zeb_2(D_\nu+i\gamma_1V_\nu)\phi,\\
\delta \psi=&F_{\mu\nu}\sigma^{\mu\nu}\zeta_2-\sqrt{2}i\sigma^\nu\zeb_1(D_\nu+i\gamma_2V_\nu)\phi,\\
\delta\bar{\la}=&-F_{\mu\nu}\zeb_1\bar{\sigma}^{\mu\nu}+\sqrt{2}i\zeta_2\sigma^\nu (D_\nu-i\gamma_1V_\nu)\bar{\phi},\\
\delta\bar{\psi}=&F_{\mu\nu}\zeb_2\bar{\sigma}^{\mu\nu}-\sqrt{2}i\zeta_1\sigma^\nu (D_\nu-i\gamma_2V_\nu)\bar{\phi},\\
\delta\phi=&-\sqrt{2}\zeta_1\psi+\sqrt{2}\zeta_2\la,\\
\delta\bar{\phi}=&-\sqrt{2}\bar{\psi}\zeb_1+\sqrt{2}\bar{\la}\zeb_2,}
for some constants $\gamma_{1,2}$, and the Killing spinor equations are as before
\eq{D_\mu\zeta_1=-iV^\rho\sigma_{\mu\rho}\zeta_1,\ D_\mu\zeta_2=iV^\rho\sigma_{\mu\rho}\zeta_2.}
We can fix the constants $\gamma_{1,2}$ from the demand that \eq{\delta_{12}F_{\mu\nu}=\delta_{\bar{1}\bar{2}}F_{\mu\nu}=0.}
We have
\eq{\onov{\sqrt{2}}\delta_{12}v_\mu&=\zeta_2\sigma^\nu\bar{\sigma}_\mu\zeta_1(D_\nu-i\gamma_1V_\nu)\bar{\phi}+\zeta_1\sigma^\nu\bar{\sigma}_\mu\zeta_2(D_\nu-i\gamma_2V_\nu)\bar{\phi}\\
&=-2\zeta_1\zeta_2\left(D_\mu-\frac{i}{2}(\gamma_1+\gamma_2)V_\mu\right)\bar{\phi}-2i(\gamma_1-\gamma_2)\zeta_1\sigma_{\mu\nu}\zeta_2V^\nu\bar{\phi}\\
&=-2D_\mu(\zeta_1\zeta_2\bar{\phi})+2\bar{\phi}D_\mu(\zeta_1\zeta_2)+i\zeta_1\zeta_2V_\mu(\gamma_1+\gamma_2)\bar{\phi}-2i(\gamma_1-\gamma_2)\zeta_1\sigma_{\mu\nu}\zeta_2V^\nu\bar{\phi}\\
&=-2D_\mu(\zeta_1\zeta_2\bar{\phi})+2i(2-\gamma_1+\gamma_2)V^\nu\zeta_1\sigma_{\mu\nu}\zeta_2\bar{\phi}+i\zeta_1\zeta_2V_\mu(\gamma_1+\gamma_2)\bar{\phi},}
where we used 
\eq{D_\mu(\zeta_1\zeta_2)=2iV^\nu\zeta_1\sigma_{\mu\nu}\zeta_2.}
The transformation is a pure gauge transformation iff
\eq{\gamma_1=-\gamma_2=1.}
The off-diagonal variations of the fermions are
\eq{\delta_{21}\la&=-2i\sigma^{\mu\nu}\zeta_1\left[\left(D_\mu+\frac{i}{2}V_\mu\right)\bar{\psi}\bar{\sigma}_\nu\zeta_2\right],\\
\delta_{12}\psi&=-2i\sigma^{\mu\nu}\zeta_2\left[\left(D_\mu-\frac{i}{2}V_\mu\right)\bar{\la}\bar{\sigma}_\nu\zeta_1\right].} 
Therefore the equations of motion are
\eq{\bar{\sigma}^\mu\left(D_\mu+\frac{i}{2}V_\mu\right)\la=\bar{\sigma}^\mu\left(D_\mu-\frac{i}{2}V_\mu\right)\psi=0.}
This is also in agreement with current conservation, and 
\eq{\delta_{1\bar{2}}\la=0}
without any other conditions.

These imply the equation of motion for the scalar
\eq{\left[D^2+V^2 \right]\phi=0,}
and the action
\eq{\lag=-\onov{4}F_{\mu\nu}F^{\mu\nu}-i\bar{\la}\bar{\sigma}^\mu\left(\nabla_\mu+\frac{i}{2}V_\mu\right)\la-i\bar{\psi}\bar{\sigma}^\mu\left(\nabla_\mu-\frac{i}{2}V_\mu\right)\psi-g^{\mu\nu}\dm\phi\dn\bar{\phi}+V^2|\phi|^2.}

The dimensions of the fermions and scalar are then
\eq{\Delta_{\la}=1\pm\left(\frac{kL}{R}-\onov{2}\right),\ \Delta_{\psi}=1\pm\left(\frac{kL}{R}+\onov{2}\right),\ \Delta_\phi=1\pm\left(\frac{kL}{R}\right).}

For $k\neq0$, the $\mathcal{N}=(2,2)$ multiplets contain a fermion of dimension $\frac{3}{2}\pm\frac{kL}{R}$, a scalar and a vector of dimension $1\pm\frac{kL}{R}$, and a fermion of dimension $\onov{2}\pm\frac{kL}{R}$.

For $k=0$, the BF bound of $\phi$ is saturated. The physical multiplet contains a $2d$ gauge field, two dimension $\frac{3}{2}$ fermions and two scalars of dimensions $1$ and $2$ (The last one comes from $v_\te$). The logarithmic scalar mode is part of the dual (non physical) multiplet and therefore the physical boundary conditions preserve the full superconformal symmetry.

\subsection{Chern Simons action from $AdS_3\times S^1$}

As we saw in section \ref{Vec}, the three dimensional action coming from the KK reduction of the four dimensional  Maxwell term gives a three dimensional Maxwell term for the $k=0$ KK mode. It is well-known \citep{marolf2006boundary} that the physical boundary conditions for a gauge field on $AdS_3$ with just a Maxwell term give a $2d$ gauge field on the boundary, as we indeed found above. On the other hand, if there is a Chern-Simons term on $AdS_3$, the physical boundary conditions will give a conserved current associated with a $U(1)$ global symmetry in the $2d$ superconformal algebra.

Therefore, it is an interesting question whether we can write a four dimensional theory that will give a Chern-Simons action on the three dimensional AdS space. One way to do this (for flat space times a circle) is to note that~\cite{Christiansen:1998xf} showed how to couple $4d$ vector and two-form multiplets in a supersymmetric and parity-odd way. Denote the superfields
\eq{\mathcal{G}&=-\onov{2}M+\frac{i}{4}\te\xi-\frac{i}{4}\bar{\te}\bar{\xi}+\onov{2}\te\sigma^\mu\bar{\te}\tilde{G}_\mu+\onov{8}\te\sigma^\mu\bar{\te}^2\dm\bar{\xi}-\onov{8}\te^2\sigma^\mu\bar{\te}\dm\xi-\onov{8}\te^2\bar{\te}^2\Box M\ ,\\
\mathcal{V}&=\te\sigma^\mu\bar{\te}A_\mu+\te^2\bar{\te}\bar{\la}+\bar{\te}^2\te\la+\te^2\bar{\te}^2\Delta\ ,\ \mathcal{W}^a=-\onov{4}\bar{D}^2D^a\mathcal{V},}
where $\tilde{G}^\mu=\onov{2}\epsilon^{\mu\nu\rho\tau}\dn B_{\rho\tau},\ B_{\rho\tau}$ is a two form, and $\mathcal{V},\ \mathcal{W}^a$ are the regular gauge field and field strength multiplets in the Wess-Zumino gauge. 
The Maxwell-Chern-Simons action in four dimensions is given by
\eq{S_{4d}=\int d^4xd^2\te\left[-\onov{8}\mathcal{W}^a\mathcal{W}_a+d^2\bar{\te}\left(-\onov{2}\mathcal{G}^2+\onov{2}m\mathcal{V}\mathcal{G}\right) \right].}
In components, the action reads
\eq{\lag_{4d}=&-\onov{4}F_{\mu\nu}F^{\mu\nu}+\onov{6}G_{\mu\nu\rho}G^{\mu\nu\rho}+m\epsilon^{\mu\nu\rho\tau}A_\mu\dn B_{\rho\tau}+2\Delta^2+\frac{i}{2}\bar{\Lambda}\Gamma^\mu\dm\Lambda\\&+\dm M\partial^\mu M+\frac{i}{4}\bar{\Xi}\Gamma^\mu\dm \Xi+im\bar{\Lambda}\Gamma_5\Xi-4mM\Delta.}
When reducing the theory on $S^1$, we get another (one form) gauge field from $w_i=B_{i\te}^{(k=0)}$. The KK zero mode then contains two vector fields $A_i,\ w_i$ with Maxwell terms and a mixed Chern Simons term with coefficient $m$. Therefore, when putting this theory on $AdS_3\times S^1$, the physical boundary conditions will lead to conserved currents and their superconformal partners as part of the two dimensional spectrum.

\bigskip
\noindent{\bf Acknowledgments}

We would like to thank Lorenzo Di Pietro, Zohar Komargodski, Ronen Plesser and Itamar Shamir for useful discussions.
This work was supported in part by an Israel Science Foundation center for excellence grant, by the I-CORE program of the Planning and Budgeting Committee and the Israel Science Foundation (grant number 1937/12), by the Minerva foundation with funding from the Federal German Ministry for Education and Research, by a Henri Gutwirth award from the Henri Gutwirth Fund for the Promotion of Research, and by the ISF within the ISF-UGC joint research program framework (grant no. 1200/14). OA is the Samuel Sebba Professorial Chair of Pure and Applied Physics.

\appendix
\section{Conventions}
\label{conventions}

\subsection{Indices}

In section \ref{SCalgebra} we use the following indices:
\begin{enumerate}
\item $\mu,\nu,\cdots$ label coordinates on the entire $d$-dimensional spacetime.
\item $a,b,\cdots$ label coordinates on $AdS_{d-q}$.
\item $A,B,\cdots$ label coordinates on $S^q$.
\item $\al,\be,\cdots$ are spinor indices.
\item $i,j,i',j',\cdots$ label the $R$-symmetry representation of the supercharges.
\end{enumerate}
In section \ref{AdS3S1} we use the following indices:

\begin{enumerate}
\item $\mu,\nu,\cdots$ label 4-dimensional $AdS_3\times S^1$ coordinates.
\item $i,j,\cdots$ label coordinates of $AdS_3$.

\item $\al,\be$ are spinor indices.
\item When written explicitly, $t,x,r,\te$ are curved $AdS_3\times S^1$ spacetime indices, and $0,1,2,3$ are flat spacetime indices.
\end{enumerate}

\subsection{Spinors and superconformal algebra}

Here we specify our conventions for the superconformal algebras and spinors in different dimensions, used in section \ref{SCalgebra}. In every dimension, we use the signature
$(-,\overbrace{+,...,+}^{d-1})$. In $4$ and $6$ dimensions, $\gamma^*$ denotes the chiral gamma matrix.

For most of the algebras, we use the conventions of \citep{Freedman:2012zz}. The only exception is the six dimensional $\mathcal{N}=(2,0)$ superconformal algebra,  for which we use \citep{Claus:1997cq}.

The four dimensional $\mathcal{N}=1$ supercharges are Majorana spinors, while the four dimensional $\mathcal{N}=2,4$ supercharges are taken to be chiral Dirac spinors, where the position of the $R$-symmetry index is used to distinguish between left and right spinors in the following way:
\eq{\gamma^*Q_i=Q_i,\ \gamma^*Q^i=-Q^i,\ \gamma^*S^i=S^i,\ \gamma^*S_i=-S_i.}

The five dimensional supercharges are symplectic Majorana spinors. In this case the $R$-symmetry indices can be lowered and raised using the anti-symmetric tensor,
\eq{X^i=\epsilon^{ij}X_j,\ X_i=X^j\epsilon_{ji}, \epsilon^{12}=\epsilon_{12}=1.}

The six dimensional supercharges are symplectic Majorana-Weyl spinors. They satisfy as in four dimensions 
\eq{\gamma^*Q_i=Q_i,\ \gamma^*Q^i=-Q^i,\ \gamma^*S^i=S^i,\ \gamma^*S_i=-S_i.}
For the $\mathcal{N}=(1,0)$ theory, the $i,j$ indices can be raised and lowered using the $\epsilon$ tensor as in five dimensions. For the $\mathcal{N}=(2,0)$ theory, the $i,j$ indices can be raised and lowered using the matrix $\Omega_{ij}$ appearing in the definition of the symplectic spinors,
\eq{X^i=\Omega^{ij}X_j,\ X_i=X^j\Omega_{ji}.}
We use a specific representation of $\Omega$
\eq{\Omega^{41}=\Omega^{23}=1, \Omega^{ij}=-\Omega^{ji}=\Omega_{ji}=-\Omega_{ij},}
and all the other components are zero.

The notations for raising and lowering spinor indices are
\eq{&\lambda^\al=C^{\al\be}\lambda_\be, \la_\al=\la^\be C_{\be\al},}
where $C$ is the charge conjugation matrix.
For its properties in the different dimensions and more details, see \cite{Freedman:2012zz, Claus:1997cq}.
  
\section{Construction of Killing spinors and supersymmetry transformations  from conformal supergravity}
\label{KSE}
From the classification made in section \ref{SCalgebra}, we find the constraints on the different superconformal transformation parameters, such that the transformations close on the desired algebra. Specifically, if we denote by $\zeta$ and $\eta$ the parameters associated with the $Q$ and $S$ transformations respectively, our choice of $\mathcal{Q}$ leads to a constraint of the form $\eta=\eta(\zeta)$.

By starting from the well-known superconformal multiplets in $4$, $5$, and $6$ dimensions, and the conformal Killing spinor equations arising in conformal supergravity, and plugging in $\eta(\zeta)$, we find the correct Killing spinor equations and transformation rules of the studied curved spacetime. We will show here explicitly how it works for $\mathcal{N}=1,2$ theories in four dimensions, but the procedure is the same for the other cases.

The conformal Killing spinor equation in $4d$ $\mathcal{N}=1$  conformal supergravity  is (see equation (16.10) in \cite{Freedman:2012zz})
\eq{\left(\partial_\mu+\onov{2}b_\mu+\onov{4}\omega_\mu^{ab}\gamma_{ab}-\frac{3i}{2}A_\mu\gamma^*\right)\zeta-\gamma_\mu\eta=0,}
where $b_\mu$ is the gauge field coupling to dilatations, $\omega_\mu^{ab}$ the spin connection and $A_\mu$ is the $U(1)_R$ gauge field.
The $AdS_3\times S^1$ solution should be obtained by plugging in $b_\mu=0$, $\eta=ic\gamma^3\gamma^*\zeta$. The equation becomes
\eq{\left(\nabla_\mu-\frac{3i}{2}A_\mu\gamma^*\right)\zeta-ic\gamma_\mu\gamma^*\zeta=0.}
By identifying \eq{2c=-V^3=\pm\onov{L},\ 3A_\mu=2A_\mu^{\rm (new\ minimal)}-V_\mu,}
we reproduce \eqref{killing} which describes $AdS_3\times S^1$ in the notations of new minimal supergravity that we used in section \ref{SCalgebra}. The other decomposition rules follow automatically. For example, the supersymmetry variation of $\Phi_\mu$, the fermionic gauge field associated with $S$ transformations, is (after plugging in $\eta(\zeta)$)
\eq{\left(\nabla_\mu-\frac{3i}{2}A_\mu\gamma^*\right)\zeta+\frac{i}{c}\gamma^3\gamma_a f^a_\mu\gamma^*\zeta,}
 where $f_\mu^a$ is the gauge field coupling to special conformal transformations. The variation vanishes if $f_\mu^a=c^2e_\mu^a$ for $\mu\neq\te$, and $f_\mu^a=-c^2e_\mu^a$ for $\mu=\te$. We see that, as expected, the specific linear combination of $Q$ and $S$ we chose gives the correct relation between the $P$ and $K$ generators in the supersymmetry algebra of section \ref{d4N1} on $AdS_3\times S^1$.

A similar computation can be done for $\mathcal{N}=2$.
The relevant part of the $\mathcal{N}=2$ superconformal Killing spinor equations is\footnote{There are several additional terms that we ignore here which are irrelevant for our discussion.  See \cite{Freedman:2012zz} for more details.}
\eq{\delta \Psi_\mu^i=\left(\nabla_\mu-\frac{i}{2}A_\mu\right)\zeta^i+U^{\ i}_{\mu\ j}\zeta^j-\gamma_\mu\eta^i=0.}

From the algebraic analysis, we know that we have two options. The first is the diagonal choice in which $\eta^i=ic\gamma^3\zeta^i$, leading to a $\mathcal{N}=(0,4)$ superconformal algebra. We will take $U^{\ i}_{\mu\ j}=0$ such that the $SU(2)_R$ symmetry is conserved, and  the Killing spinor equations become
\eq{\left(\nabla_\mu-\frac{i}{2}A_\mu\right)\zeta^i-ic\gamma_\mu\gamma^3\zeta^i=0.}
We can identify as before $c=\pm\onov{2L}$ and redefine $A'_\mu=\onov{2}A_\mu+ce_{\mu 3}=\onov{2}(A_\mu-V_\mu)$ such that the equations become
\eq{D_\mu\zeta^i=(\nabla_\mu-iA'_\mu)\zeta^i=-iV^\rho\sigma_{\mu\rho}\zeta^i.}

The second option is the twisted one in which $\eta^1=ic\gamma^3\zeta^2,\ \eta^2=ic\gamma^3\zeta^1$, leading to a $\mathcal{N}=(2,2)$ superconformal algebra. If we take $U^{\ i}_{\mu\ j}=0$, the Killing spinor equations are
\eq{&\left(\nabla_\mu-\frac{i}{2}A_\mu\right)\zeta^1-ic\gamma_\mu\gamma_3\zeta^2=0,\\
&\left(\nabla_\mu-\frac{i}{2}A_\mu\right)\zeta^2-ic\gamma_\mu\gamma_3\zeta^1=0.}
In the basis $\zeta^\pm=\onov{\sqrt{2}}(\zeta^1\pm\zeta^2)$ the equations take the form
\eq{\left(\nabla_\mu-\frac{i}{2}A_\mu\right)\zeta^\pm=\pm ice_{\mu 3}\zeta^\pm\pm2ic\sigma_{\mu 3}\zeta^{\pm}.}
By the same identification as before, we find
\eq{D_\mu\zeta^{\pm}=\mp i\left(\onov{2}V_\mu+V^{\rho}\sigma_{\mu\rho}\right)\zeta^{\pm}.}
The solutions to these equations admit a relative phase of $e^{\frac{iR}{L}\te}$ between the spinors, and is therefore supersymmetric only if the radii satisfy the quantization condition $\frac{R}{L}\in \mathbb{N}$. The reason is that as explained in appendix \ref{Rgauge}, the supersymmetry transformations are consistent only if the Killing spinors are single valued. In that case, $\zeta^{\pm}$ can both be single valued at the same time only if the relative phase between them is an integer multiple of $\te$. If we want the theory to be supersymmetric for general radii, we can turn on a specific $U^{\ i}_{\mu j}$ and eliminate the relative phase between the Killing spinors. By doing so the Killing spinor equations take the form
\eq{D_\mu\zeta^{\pm}=\mp iV^\rho\sigma_{\mu\rho}\zeta^{\pm}.}

In the same way we construct the supersymmetric multiplets from the superconformal multiplets. For example, we can take a superconformal hypermultiplet with the transformation rules
\eq{&\delta\mat{\phi_1\\\bar{\phi}_2}=-\sqrt{2}\psi_1\mat{\zeta_1\\\zeta_2}-\sqrt{2}\bar{\psi}_2\mat{-\zeb_2\\\zeb_1},\\
&\delta\psi_1=-i\sqrt{2}\sigma^\nu\mat{\zeb_1&\zeb_2}D_\nu\mat{\phi_1\\\bar{\phi}_2}-2\sqrt{2}i\mat{\bar{\eta}_1&\bar{\eta}_2}\mat{\phi_1\\\bar{\phi}_2},\\
&\delta\bar{\psi}_2=i\sqrt{2}\bar{\sigma}^\nu\mat{-\zeta_2&\zeta_1}D_\nu\mat{\phi_1\\\bar{\phi}_2}-2\sqrt{2}i\mat{-\eta_2&\eta_1}\mat{\phi_1\\\bar{\phi}_2},}
and plug in the diagonal decomposition rule
\eq{\eta^i=-\frac{i}{2}V_\mu\bar{\sigma}^\mu\zeta^i.}
We get the supersymmetry transformation rules on $AdS_3\times S^1$
  \eq{&\delta\psi_1=-i\sqrt{2}\sigma^\nu\mat{\zeb_1&\zeb_2}D_\nu\mat{\phi_1\\\bar{\phi}_2}-\sqrt{2}\sigma^\nu\mat{\bar{\zeta}_1&\bar{\zeta}_2}V_\nu\mat{\phi_1\\\bar{\phi}_2},\\
&\delta\bar{\psi}_2=i\sqrt{2}\bar{\sigma}^\nu\mat{-\zeta_2&\zeta_1}D_\nu\mat{\phi_1\\\bar{\phi}_2}-\sqrt{2}\bar{\sigma}^\nu\mat{-\zeta_2&\zeta_1}V_\nu\mat{\phi_1\\\bar{\phi}_2}.}
In the same way, this is done for the other multiplets and the other supersymmetries.

\section{Comments about R-gauge transformations}\label{Rgauge}

As we saw in section \ref{AdS3S1}, there is a freedom in choosing the background R-symmetry gauge field, which results in the Killing spinor in a phase $\zeta\sim e^{in\te}$. Above we always chose $n=0$, but let us see what happens in the more general case. In the four dimensional $\mathcal{N}=1$ case on $AdS_3\times S^1$, is we use such a more general background $A_{\mu}$ and represent some scalar field with $4d$ $R$-charge $\hat{q}$ using its KK modes around the $S^1$, the $U(1)_R$ charge of the $k$'th KK mode in the two dimensional superconformal algebra is $\left(\frac{kL}{R}+\hat{q}+\frac{n\hat{q}L}{R}\right)$ (this can be seen from \eqref{identification}).
We wish to understand if different choices for $n$ lead to different physical theories.
\begin{itemize}
\item If $\hat{q}$ is an integer, $n$ can always be absorbed by shifting the definition of $k$, and therefore it has no consequences on the physics. \\
\item If $\hat{q}$ is irrational, different choices of $n$ will lead to a different spectrum of the theory with different dimensions and two-dimensional $R$-charges, and thus to a different theory on $AdS_3\times S^1$. \\
\item If $\hat{q}=\frac{M}{N}$ is rational (where $M$ and $N$ are coprime integers), there are $N$ different physical theories with different spectra that can be obtained by changing $n$, while two theories with $n$ and $n'=n+N$ are physically equivalent.
\end{itemize}

The Killing spinors should clearly be periodic in $\te$ for the supersymmetry generators to be well-defined. In particular,
if the Killing spinors have a phase $\zeta\sim e^{in\te}$, the supersymmetry transformations relate different KK modes with $k_{fermion}-k_{boson}=\pm n$, which exist only if $n$ is an integer. 

\section{Six dimensional superconformal algebras on $AdS_6$ and $AdS_4\times S^2$}\label{6d}

We claim that the six dimensional $\mathcal{N}=(1,0)$ and $\mathcal{N}=(2,0)$ superconformal (SC) algebras do not have subalgebras that close on the isometries of $AdS_6$ and $AdS_4\times S^2$. One argument for this (say in the $AdS_6$ case) is that $AdS_6$ is conformally related to flat space with a boundary, and the boundary conditions on a spinor necessarily modify its chirality, which cannot be done in a supersymmetric way when the supercharges are chiral.

We can also see this in a purely algebraic way. The six dimensional algebra on $AdS_6$ must be equivalent to the $F(4)$ SC algebra in 5 dimensions which is the only 5 dimensional superconformal algebra. This algebra includes 16 supercharges and $SU(2)$ as its $R$-symmetry. The $\mathcal{N}=(2,0)$ 6 dimensional superconformal algebra has 32 supercharges, so the numbers fit (after breaking half) but there is no choice of supercharges that will break the $R$-symmetry to $SU(2)$ (but to a larger group). If we discuss $AdS_4\times S^2$, then by counting supercharges, the 6 dimensional $\mathcal{N}=(1,0),\ (2,0)$ should correspond to the 3 dimensional $\mathcal{N}=2,\ 4$ superconformal algebras with $R$-symmetries of $SO(2),\ SO(4)$ respectively.
These $R$-symmetry groups must be a mixture of the $SO(3)\ S^2$ isometries and a subgroup of the six dimensional $R$-symmetry group. For the two cases this cannot be done.
More explicitly, in order to form such a subalgebra, there is a limited amount of options to connect the supercharges $Q_{i}$ and $S_{j}$. Specifically, for $AdS_6$, we can choose either $Q_i+icS^{i'}$ or $Q_i+ic\gamma^*S^{i'}$. In both cases, the anti-commutators of these charges contain the dilatation operator $D$, which does not keep us in the bosonic subalgebra we want, but rather brings us back to the full superconformal algebra. The same arguments apply in the case of $AdS_4\times S^2$, with the options $Q_i+ic\gamma^{45}S^{i'}$ and  $Q_i+ic\gamma^*\gamma^{45}S^{i'}$.

\section{Super-Virasoro algebra on $AdS_3\times S^{d-3}$}

In the analysis of section \ref{SCalgebra} we found the possible $2d$ superconformal algebras arising in field theories on $AdS_3\times S^{d-3}$. The conformal generators that are dual to the $AdS$ isometries are just the global ones (which in the language of Virasoro generators are denoted by ${\cal L}_{\pm1,0}$, $\bar{\cal L}_{\pm1,0}$), and the superconformal algebras we found contain just the global superconformal algebra and not the full super-Virasoro algebra. As was shown in \citep{brown1986central}, when there is a fluctuating gravity theory on asymptotically $AdS_3$ spacetimes, the symmetry group  includes the entire Virasoro group. Here we studied curved manifolds without discussing gravity at all, and we can ask whether our field theories can come from some $G_N\rightarrow 0$ limit of a gravitational theory.

One basic requirement is that the superconformal algebras we found can be extended into a super-Virasoro algebra. We encounter problems in two cases, the $\mathcal{N}=(0,4)$ and $\mathcal{N}=(4,4)$ superconformal algebras that come from $4d$ $\mathcal{N}=2,4$ supersymmetric field theories on $AdS_3\times S^1$. These algebras contain a central extension beyond the known superconformal algebras. As far as we know, these central charges are inconsistent with the extension to a super-Virasoro algebra. Thus, we claim that such field theories cannot appear as decoupled sectors of gravitational theories on $AdS_3$, though they could arise as decoupled sectors of gravitational theories in higher dimensions (e.g. on D3-branes wrapping $AdS_3\times S^1$ inside $AdS_5\times S^5$). It
would be interesting to understand how this is consistent with string theories on $AdS_3\times S^3\times T^4$, which give rise to such super-Virasoro algebras.

\section{Summary of maximal supersymmetry algebras}
\label{Summary}
\begin{tabular}{c|c|c|c|c|c}
Dimension&    $\mathcal{N}$ &supercharges & spacetime & superconformal algebra &  R-symmetry \\
3  &   N &  2N  & $AdS_3$ & $2d$ $\mathcal{N}=(n,N-n)$& $SO(n)\times SO(N-n)$\\
4  &   1 &  4 & $AdS_4$ & $3d$ $\mathcal{N}=1$ &  -   \\
4  &   2 &  8 & $AdS_4$ & $3d$ $\mathcal{N}=2$ &  SO(2)   \\
4  &   4 &  16 & $AdS_4$ & $3d$ $\mathcal{N}=4$ & SO(4)    \\
4  &   1 &  4 & $AdS_3\times S^1$ & $2d$ $\mathcal{N}=(0,2)$ & U(1)    \\
4  &   2 &  8 & $AdS_3\times S^1$ & $2d$ $\mathcal{N}=(0,4)$+central &  $SU(2)$   \\
4  &   2 &  8 & $AdS_3\times S^1$ & $2d$ $\mathcal{N}=(2,2)$& $U(1)\times U(1)$    \\
4  &   4 &  16 & $AdS_3\times S^1$ &$2d$ $\mathcal{N}=(0,8)$ & U(4)    \\
4  &   4 &  16 & $AdS_3\times S^1$ & $2d$ $\mathcal{N}=(2,6)$ &  $U(1)\times U(3)$   \\
4  &   4 &  16 & $AdS_3\times S^1$ & $2d$ $\mathcal{N}=(4,4)$+central &   $SU(2)\times SU(2)$  \\
5  &   F(4) &  8 & $AdS_5$ & $4d$ $\mathcal{N}=1$& U(1)    \\
5  &  F(4) &  8 & $AdS_4\times S^1$ & $3d$ $\mathcal{N}=2$& SO(2)   \\
5  &  F(4) &  8 & $AdS_3\times S^2$ & $2d$ `large' $\mathcal{N}=(0,4)$ & $SU(2)\times SU(2)$   \\
6  &  (0,1) &  8 & $AdS_6$ & none & - \\
6  &  (0,2) &  16 & $AdS_6$ & none & - \\
6  &  (0,1) &  8 & $AdS_5\times S^1$ &$4d$ $\mathcal{N}=1$& U(1) \\
6  &  (0,2) &  16 & $AdS_5\times S^1$ & $4d$ $\mathcal{N}=2$& SU(2) \\
6  &  (0,1) &  8 & $AdS_4\times S^2$ & none & - \\
6  &  (0,2) &  16 & $AdS_4\times S^2$ & none & - \\
6  &  (0,1) &  8 & $AdS_3\times S^3$ & $2d$ `large' $\mathcal{N}=(0,4)$&$SU(2)\times SU(2)$ \\
6  &  (0,2) &  16 & $AdS_3\times S^3$ & $2d$ $\mathcal{N}=(0,8)$ & $Spin(4)\times SU(2)$\\
6  &  (0,2) &  16 & $AdS_3\times S^3$ & $2d$ `large' $\mathcal{N}=(4,4)$ & $SO(4)\times SO(4)$\\
\end{tabular}

\bibliography{bibclassification}

\providecommand{\href}[2]{#2}\begingroup\raggedright\begin{thebibliography}{10}

\bibitem{Pestun:2007rz}
V.~Pestun, {\it {Localization of gauge theory on a four-sphere and
  supersymmetric Wilson loops}},  {\em Commun. Math. Phys.} {\bf 313} (2012)
  71--129, [\href{http://arxiv.org/abs/0712.2824}{{\tt arXiv:0712.2824}}].

\bibitem{Aharony:2015zea}
O.~Aharony, M.~Berkooz, and S.-J. Rey, {\it {Rigid holography and
  six-dimensional $ \mathcal{N}=\left(2,0\right) $ theories on $AdS_{5}\times
  \mathbb{S}^{1}$}},  {\em JHEP} {\bf 03} (2015) 121,
  [\href{http://arxiv.org/abs/1501.02904}{{\tt arXiv:1501.02904}}].

\bibitem{Zumino:1977av}
B.~Zumino, {\it {Nonlinear Realization of Supersymmetry in de Sitter Space}},
  {\em Nucl. Phys.} {\bf B127} (1977) 189.

\bibitem{Ivanov:1979ft}
E.~A. Ivanov and A.~S. Sorin, {\it {{Wess-Zumino} Model as Linear Sigma Model
  of Spontaneously Broken Conformal and Osp(1,4) Supersymmetries}},  {\em Sov.
  J. Nucl. Phys.} {\bf 30} (1979) 440. [Yad. Fiz.30,853(1979)].

\bibitem{Ivanov:1980vb}
E.~A. Ivanov and A.~S. Sorin, {\it {Superfield formulation of OSp(1,4)
  supersymmetry}},  {\em J. Phys.} {\bf A13} (1980) 1159--1188.

\bibitem{Sakai:1984nc}
N.~Sakai and Y.~Tanii, {\it {Supersymmetry and Vacuum Energy in Anti-de Sitter
  Space}},  {\em Phys. Lett.} {\bf B146} (1984) 38.

\bibitem{Burgess:1984rz}
C.~P. Burgess, {\it {Supersymmetry Breaking and Vacuum Energy on Anti-de Sitter
  Space}},  {\em Nucl. Phys.} {\bf B259} (1985) 473.

\bibitem{Burgess:1984ti}
C.~P. Burgess and C.~A. Lutken, {\it {Propagators and Effective Potentials in
  Anti-de Sitter Space}},  {\em Phys. Lett.} {\bf B153} (1985) 137.

\bibitem{Burges:1985qq}
C.~J.~C. Burges, D.~Z. Freedman, S.~Davis, and G.~W. Gibbons, {\it
  {Supersymmetry in Anti-de Sitter Space}},  {\em Annals Phys.} {\bf 167}
  (1986) 285.

\bibitem{Adams:2011vw}
A.~Adams, H.~Jockers, V.~Kumar, and J.~M. Lapan, {\it {N=1 Sigma Models in
  $AdS_4$}},  {\em JHEP} {\bf 12} (2011) 042,
  [\href{http://arxiv.org/abs/1104.3155}{{\tt arXiv:1104.3155}}].

\bibitem{Aharony:2010ay}
O.~Aharony, D.~Marolf, and M.~Rangamani, {\it {Conformal field theories in
  anti-de Sitter space}},  {\em JHEP} {\bf 02} (2011) 041,
  [\href{http://arxiv.org/abs/1011.6144}{{\tt arXiv:1011.6144}}].

\bibitem{Aharony:2012jf}
O.~Aharony, M.~Berkooz, D.~Tong, and S.~Yankielowicz, {\it {Confinement in
  Anti-de Sitter Space}},  {\em JHEP} {\bf 02} (2013) 076,
  [\href{http://arxiv.org/abs/1210.5195}{{\tt arXiv:1210.5195}}].

\bibitem{Kuzenko:2007aj}
S.~M. Kuzenko and G.~Tartaglino-Mazzucchelli, {\it {Five-dimensional N = 1 AdS
  superspace: Geometry, off-shell multiplets and dynamics}},  {\em Nucl. Phys.}
  {\bf B785} (2007) 34--73, [\href{http://arxiv.org/abs/0704.1185}{{\tt
  arXiv:0704.1185}}].

\bibitem{Kuzenko:2008kw}
S.~M. Kuzenko and G.~Tartaglino-Mazzucchelli, {\it {Conformally flat
  supergeometry in five dimensions}},  {\em JHEP} {\bf 06} (2008) 097,
  [\href{http://arxiv.org/abs/0804.1219}{{\tt arXiv:0804.1219}}].

\bibitem{Kuzenko:2008qw}
S.~M. Kuzenko and G.~Tartaglino-Mazzucchelli, {\it {Field theory in 4D N=2
  conformally flat superspace}},  {\em JHEP} {\bf 10} (2008) 001,
  [\href{http://arxiv.org/abs/0807.3368}{{\tt arXiv:0807.3368}}].

\bibitem{Kuzenko:2011rd}
S.~M. Kuzenko and G.~Tartaglino-Mazzucchelli, {\it {Three-dimensional N=2 (AdS)
  supergravity and associated supercurrents}},  {\em JHEP} {\bf 12} (2011) 052,
  [\href{http://arxiv.org/abs/1109.0496}{{\tt arXiv:1109.0496}}].

\bibitem{Butter:2012jj}
D.~Butter, S.~M. Kuzenko, U.~Lindstrom, and G.~Tartaglino-Mazzucchelli, {\it
  {Extended supersymmetric sigma models in $AdS_4$ from projective
  superspace}},  {\em JHEP} {\bf 05} (2012) 138,
  [\href{http://arxiv.org/abs/1203.5001}{{\tt arXiv:1203.5001}}].

\bibitem{Kuzenko:2012bc}
S.~M. Kuzenko, U.~Lindstrom, and G.~Tartaglino-Mazzucchelli, {\it
  {Three-dimensional (p,q) AdS superspaces and matter couplings}},  {\em JHEP}
  {\bf 08} (2012) 024, [\href{http://arxiv.org/abs/1205.4622}{{\tt
  arXiv:1205.4622}}].

\bibitem{Butter:2012vm}
D.~Butter, S.~M. Kuzenko, and G.~Tartaglino-Mazzucchelli, {\it {Nonlinear sigma
  models with AdS supersymmetry in three dimensions}},  {\em JHEP} {\bf 02}
  (2013) 121, [\href{http://arxiv.org/abs/1210.5906}{{\tt arXiv:1210.5906}}].

\bibitem{Kuzenko:2013uya}
S.~M. Kuzenko, U.~Lindstrom, M.~Rocek, I.~Sachs, and
  G.~Tartaglino-Mazzucchelli, {\it {Three-dimensional $\mathcal{N} =$ 2
  supergravity theories: From superspace to components}},  {\em Phys. Rev.}
  {\bf D89} (2014), no.~8 085028, [\href{http://arxiv.org/abs/1312.4267}{{\tt
  arXiv:1312.4267}}].

\bibitem{Kuzenko:2014mva}
S.~M. Kuzenko and G.~Tartaglino-Mazzucchelli, {\it {N = 4 supersymmetric
  Yang-Mills theories in $AdS_3$}},  {\em JHEP} {\bf 05} (2014) 018,
  [\href{http://arxiv.org/abs/1402.3961}{{\tt arXiv:1402.3961}}].

\bibitem{Kuzenko:2014eqa}
S.~M. Kuzenko, J.~Novak, and G.~Tartaglino-Mazzucchelli, {\it {Symmetries of
  curved superspace in five dimensions}},  {\em JHEP} {\bf 10} (2014) 175,
  [\href{http://arxiv.org/abs/1406.0727}{{\tt arXiv:1406.0727}}].

\bibitem{Shuster:1999zf}
E.~Shuster, {\it {Killing spinors and supersymmetry on AdS}},  {\em Nucl.
  Phys.} {\bf B554} (1999) 198--214,
  [\href{http://arxiv.org/abs/hep-th/9902129}{{\tt hep-th/9902129}}].

\bibitem{Bandos:2002nn}
I.~A. Bandos, E.~Ivanov, J.~Lukierski, and D.~Sorokin, {\it {On the
  superconformal flatness of AdS superspaces}},  {\em JHEP} {\bf 06} (2002)
  040, [\href{http://arxiv.org/abs/hep-th/0205104}{{\tt hep-th/0205104}}].

\bibitem{Festuccia:2011ws}
G.~Festuccia and N.~Seiberg, {\it {Rigid Supersymmetric Theories in Curved
  Superspace}},  {\em JHEP} {\bf 06} (2011) 114,
  [\href{http://arxiv.org/abs/1105.0689}{{\tt arXiv:1105.0689}}].

\bibitem{Gaiotto:2008sa}
D.~Gaiotto and E.~Witten, {\it {Supersymmetric Boundary Conditions in N=4 Super
  Yang-Mills Theory}},  {\em J. Statist. Phys.} {\bf 135} (2009) 789--855,
  [\href{http://arxiv.org/abs/0804.2902}{{\tt arXiv:0804.2902}}].

\bibitem{Dumitrescu:2012ha}
T.~T. Dumitrescu, G.~Festuccia, and N.~Seiberg, {\it {Exploring Curved
  Superspace}},  {\em JHEP} {\bf 08} (2012) 141,
  [\href{http://arxiv.org/abs/1205.1115}{{\tt arXiv:1205.1115}}].

\bibitem{Nahm:1977tg}
W.~Nahm, {\it {Supersymmetries and their Representations}},  {\em Nucl. Phys.}
  {\bf B135} (1978) 149.

\bibitem{Freedman:2012zz}
D.~Z. Freedman and A.~Van~Proeyen, {\it {Supergravity}}, .

\bibitem{Wess:1992cp}
J.~Wess and J.~Bagger, {\em {Supersymmetry and supergravity}}.
\newblock 1992.

\bibitem{Butter:2015tra}
D.~Butter, G.~Inverso, and I.~Lodato, {\it {Rigid 4D N=2 supersymmetric
  backgrounds and actions}},  \href{http://arxiv.org/abs/1505.03500}{{\tt
  arXiv:1505.03500}}.

\bibitem{Bergshoeff:2001hc}
E.~Bergshoeff, S.~Cucu, M.~Derix, T.~de~Wit, R.~Halbersma, and A.~Van~Proeyen,
  {\it {Weyl multiplets of N=2 conformal supergravity in five-dimensions}},
  {\em JHEP} {\bf 06} (2001) 051,
  [\href{http://arxiv.org/abs/hep-th/0104113}{{\tt hep-th/0104113}}].

\bibitem{Park:1999cw}
J.-H. Park, {\it {Superconformal symmetry in three-dimensions}},  {\em J. Math.
  Phys.} {\bf 41} (2000) 7129--7161,
  [\href{http://arxiv.org/abs/hep-th/9910199}{{\tt hep-th/9910199}}].

\bibitem{Coomans:2011ih}
F.~Coomans and A.~Van~Proeyen, {\it {Off-shell N=(1,0), D=6 supergravity from
  superconformal methods}},  {\em JHEP} {\bf 02} (2011) 049,
  [\href{http://arxiv.org/abs/1101.2403}{{\tt arXiv:1101.2403}}]. [Erratum:
  JHEP01,119(2012)].

\bibitem{Ali:1999ut}
A.~Ali, {\it {Classification of two-dimensional N=4 superconformal
  symmetries}},  \href{http://arxiv.org/abs/hep-th/9906096}{{\tt
  hep-th/9906096}}.

\bibitem{Fradkin:1992bz}
E.~S. Fradkin and V.~{\relax Ya}. Linetsky, {\it {Results of the classification
  of superconformal algebras in two-dimensions}},  {\em Phys. Lett.} {\bf B282}
  (1992) 352--356, [\href{http://arxiv.org/abs/hep-th/9203045}{{\tt
  hep-th/9203045}}].

\bibitem{Sohnius:1981tp}
M.~F. Sohnius and P.~C. West, {\it {An Alternative Minimal Off-Shell Version of
  N=1 Supergravity}},  {\em Phys.Lett.} {\bf B105} (1981) 353.

\bibitem{marolf2006boundary}
D.~Marolf and S.~F. Ross, {\it {Boundary Conditions and New Dualities: Vector
  Fields in AdS/CFT}},  {\em JHEP} {\bf 11} (2006) 085,
  [\href{http://arxiv.org/abs/hep-th/0606113}{{\tt hep-th/0606113}}].

\bibitem{breitenlohner1982stability}
P.~Breitenlohner and D.~Z. Freedman, {\it {Stability in Gauged Extended
  Supergravity}},  {\em Annals Phys.} {\bf 144} (1982) 249.

\bibitem{breitenlohner1982positive}
P.~Breitenlohner and D.~Z. Freedman, {\it {Positive Energy in anti-De Sitter
  Backgrounds and Gauged Extended Supergravity}},  {\em Phys. Lett.} {\bf B115}
  (1982) 197.

\bibitem{Christiansen:1998xf}
H.~R. Christiansen, M.~S. Cunha, J.~A. Helayel-Neto, L.~R.~U. Manssur, and
  A.~L. M.~A. Nogueira, {\it {N=2 Maxwell-Chern-Simons model with anomalous
  magnetic moment coupling via dimensional reduction}},  {\em Int. J. Mod.
  Phys.} {\bf A14} (1999) 147--160,
  [\href{http://arxiv.org/abs/hep-th/9802096}{{\tt hep-th/9802096}}].

\bibitem{Claus:1997cq}
P.~Claus, R.~Kallosh, and A.~Van~Proeyen, {\it {M five-brane and superconformal
  (0,2) tensor multiplet in six-dimensions}},  {\em Nucl. Phys.} {\bf B518}
  (1998) 117--150, [\href{http://arxiv.org/abs/hep-th/9711161}{{\tt
  hep-th/9711161}}].

\bibitem{brown1986central}
J.~D. Brown and M.~Henneaux, {\it {Central Charges in the Canonical Realization
  of Asymptotic Symmetries: An Example from Three-Dimensional Gravity}},  {\em
  Commun. Math. Phys.} {\bf 104} (1986) 207--226.

\end{thebibliography}\endgroup

\end{document}